\documentclass[12pt]{article}
 \pdfoutput=1
\usepackage[centertags]{amsmath}
\usepackage{hyperref}
\usepackage{color}
\usepackage{amsfonts}
\usepackage{amssymb}
\usepackage{amsthm}
\usepackage{cite}
\usepackage{fullpage}
\usepackage{graphicx}
\usepackage{pgf,pgfarrows,pgfnodes}
\usepackage{graphics}

\let\a=u

       \let\C=\Chi    
\newcommand{\Z}{{\mathbb Z}}
\newcommand{\R}{{\mathbb R}}
\newcommand{\C}{{\mathbb C}}
\newcommand{\be}{\begin{equation}}
\newcommand{\eeq}{\end{equation}}
\newcommand{\bea}{\begin{eqnarray}}
\newcommand{\eea}{\end{eqnarray}}
\newcommand{\ba}{\begin{array}}
\newcommand{\ea}{\end{array}}
\def\nn{\nonumber}
\newcommand{\ft}[2]{{\textstyle\frac{#1}{#2}}}

\newcommand{\ee}{\end{equation} }

\newcommand{\one}{{\rm 1\kern -.9mm l}}

\newdimen\tableauside\tableauside=1.0ex
\newdimen\tableaurule\tableaurule=0.4pt
\newdimen\tableaustep

\def\phantomhrule#1{\hbox{\vbox to0pt{\hrule height\tableaurule
width#1\vss}}}
\def\phantomvrule#1{\vbox{\hbox to0pt{\vrule width\tableaurule
height#1\hss}}}
\def\sqr{\vbox{%
 \phantomhrule\tableaustep
\hbox{\phantomvrule\tableaustep\kern\tableaustep\phantomvrule\tableaustep}%
 \hbox{\vbox{\phantomhrule\tableauside}\kern-\tableaurule}}}
\def\squares#1{\hbox{\count0=#1\noindent\loop\sqr
 \advance\count0 by-1 \ifnum\count0>0\repeat}}
\def\tableau#1{\vcenter{\offinterlineskip
 \tableaustep=\tableauside\advance\tableaustep by-\tableaurule
 \kern\normallineskip\hbox
   {\kern\normallineskip\vbox
     {\gettableau#1 0 }%
    \kern\normallineskip\kern\tableaurule}%
 \kern\normallineskip\kern\tableaurule}}
\def\gettableau#1 {\ifnum#1=0\let\next=\null\else
 \squares{#1}\let\next=\gettableau\fi\next}
\newcommand{\Yfund}{\tableau{1}}

\numberwithin{equation}{section}

\begin{document}

\font\cmss=cmss10 \font\cmsss=cmss10 at 7pt

\vskip -0.5cm
%\rightline{\small{\tt MPP-2007-162}}
\rightline{\small{\tt ROM2F/2013/05}}
\rightline{\small{\tt DFPD-13/TH/09}}

\vskip .7 cm
%\hfill IC/2004/ \vskip .1in \hfill CPHT \vskip .1in \hfill hep-th/yymmnnn

\hfill
\vspace{18pt}
\begin{center}
{\Large \textbf{Exact results in $\mathcal{N}=2$ gauge theories  }}
\end{center}

\vspace{6pt}
\begin{center}

  {\textsl{Francesco Fucito$^{\dagger}$ \footnote{\scriptsize \tt francesco.fucito@roma2.infn.it}, Jose Francisco Morales $^{\dagger }$ \footnote{\scriptsize \tt francisco.morales@roma2.infn.it}, Rubik Poghossian $^{\sharp }$ \footnote{\scriptsize \tt poghos@yerphi.am}  \\ \& Daniel Ricci Pacifici $^\ddagger$ \footnote{\scriptsize \tt daniel.riccipacifici@pd.infn.it}}}

\vspace{1cm}
%\textit{\small $^a$ Max-Planck-Institut f\"{u}r Physik -- Theorie,\\
 %                   F\"{o}hringer Ring 6,  D-80805 M\"{u}nchen, Germany}\\ \vspace{6pt}
$\dagger$ \textit{\small I.N.F.N. Sezione di Roma ``TorVergata'' \&\\  Dipartimento di Fisica, Universit\`a di Roma ``TorVergata", \\
Via della Ricerca Scientifica, 00133 Roma, Italy }\\  \vspace{6pt}
$\sharp$ \textit{\small Yerevan Physics Institute,\\
Alikhanian Br. 2, 0036 Yerevan, Armenia}\\  \vspace{6pt}
$ \ddagger $ \textit{\small Dipartimento di Fisica e Astronomia, Universit\`a degli Studi di Padova \&\\  I.N.F.N, Sezione di Padova,\\
Via Marzolo 8, 35131, Padova, Italy}\\  \vspace{6pt}

\end{center}

\begin{center}
\textbf{Abstract}
\end{center}

\vspace{4pt} {\small

\noindent We derive exact formulae for the partition function and the expectation values of Wilson/'t Hooft loops, thus directly checking their S-duality transformations. We focus on 
a special class of ${\cal N}=2$ gauge theories on $S^4$ with fundamental matter. In particular we show that, for a specific choice of
the masses, the matrix model integral defining the gauge theory partition function localizes
 around a finite set of critical points where it can be explicitly evaluated and written in terms of generalized hypergeometric functions.  
 From the AGT perspective the gauge theory partition function, evaluated with this choice of masses, is viewed as a four point correlator 
  involving the  insertion of a degenerated field. The well known simplicity of the degenerated correlator reflects the fact that for 
these choices of masses  only a very restrictive type of instanton configurations contributes to the gauge theory partition function.}

\newpage

\tableofcontents

\section{Introduction}
 The study of Wilson loops in supersymmetric gauge theories has received a lot of attention in recent years.
According to holography, the expectation value of a supersymmetric Wilson loop in ${\cal N}=4$ gauge theory at strong coupling is
computed by the area of the minimal surface on AdS swept by an open string ending on the Wilson loop  itself \cite{Maldacena:1998im}.
 This result is confirmed  by a perturbative computation in ${\cal N}=4$  \cite{Erickson:2000af}.
 In fact the contribution of each relevant Feynman integral is shown to be one and the multiplicities of the Feynman diagrams are evaluated by a
gaussian matrix model. For circular Wilson loops the matrix model integral  can be computed and it gives  a Bessel function reproducing  the right
strong coupling asymptotics predicted by gravity.  This formula was rigorously proven in \cite{Pestun:2007rz} using localization techniques.
The results were also extended to ${\cal N}=2$, where the measure of the matrix model integral is completed by  the one-loop,  $Z_{\rm one-loop}$,
and instanton contributions, $Z_{\rm inst}$, to the partition function of the gauge theory. There are two complications in extending the exact
${\cal N}=4$ result to ${\cal N}=2$. First,  $Z_{\rm inst}$ can be computed only order by order in $q=e^{2 \pi i \tau}$. Second,
for a general form of the perturbative and non perturbative contributions the matrix model integral is not amenable to an analytic treatment.
In this paper we show that for a special (but not very restrictive)  choice of the gauge theory parameters
  both difficulties can be overcome and exact formulae can be derived.

 We focus on ${\cal N}=2$ gauge theories with gauge group $U(N)$ and matter in the fundamental and anti-fundamental representations.
 The use of localization requires that the gauge theory be placed  on  a non-trivial $\epsilon$-background lifting its Lorentz symmetries.
From a physical point of view the introduction of the $\epsilon$ background can be viewed as a gravitational
$\Omega$-background \cite{Losev:2003py}  or as the result of type IIB RR fluxes \cite{Billo:2006jm,Lambert:2013lxa,Hellerman:2012rd}.
    We will consider both the cases of gauge theories on $\R^4$ and $S^4$. In the former case  one should also specify the expectation
values  $a\in SU(N)$ of the scalar field at infinity.  The partition function, $Z(a)$, depends then
   on the masses $m$, vevs $a$, $\epsilon_\ell$ deformations and the gauge coupling $q$ and it can be computed order by order in $q$ via localization techniques  \cite{Nekrasov:2002qd,Flume:2002az,Bruzzo:2002xf}.
 % (see   \cite{Moore:1998et} for earlier applications of  localization ideas).
 In  the limit where both $\epsilon_\ell$ are small,
 it becomes  $Z(a)\approx e^{-{1\over \epsilon_1 \epsilon_2} {\cal F}_{\rm SW}}$, where
 ${\cal F}_{\rm SW}$ is the Seiberg-Witten prepotential  determining the two-derivative effective action of the gauge theory.
 On the other hand, in the limit when $\epsilon_1$ is small but $\epsilon_2=\epsilon$ is finite, the $\epsilon$-deformed dynamics can be
shown to be in correspondence with that of certain quantum integrable systems with $\epsilon$ playing the role of the Planck constant  \cite{Nekrasov:2009uh,Nekrasov:2009ui,Nekrasov:2009rc}.
Both limits can be  treated via saddle point techniques leading to a Seiberg-Witten curve (or its $\epsilon$-deformed
version) \cite{Nekrasov:2003rj,Poghossian:2010pn,Fucito:2011pn} \footnote{The generalisation of this analysis to ADE quiver gauge theories can be found in \cite{Nekrasov:2012xe,Fucito:2012xc}.}.

   In this paper we consider the case where both $\epsilon_1$ and $\epsilon_2$ are finite.
 Drawing inspiration from previous works in CFT's, we show that the masses can be chosen in such a way that only a
very peculiar class of gauge  instanton configurations can contribute to $Z$ and that the full instanton sum can be explicitly evaluated
(see \cite{Bonelli:2011wx,Bonelli:2011aa} for previous works in this direction).  In the AGT dual description of the
theory  \cite{Alday:2009aq}, where the gauge theory instanton partition function is described by a four-point conformal block of
the Toda field theory, this specific choice of mass corresponds to the insertion of a degenerated field. The conformal block is then determined
by a differential equation that, in the simplest case, can be solved in terms of generalized hypergeometric
functions \cite{Zamolodchikov:1995aa,Fateev:2005gs,Fateev:2007ab}.

      In the case of a gauge theory  on $S^4$ the partition function  is given by the integral $\int da |Z(a)|^2$
 with $Z$ and $\bar Z$ the contributions of the instantons and  anti-instantons  located at the north and south poles of the
two charts in $S^4$ \cite{Pestun:2007rz,Alday:2009fs}.
     The partition function is computed for $\epsilon_1=\epsilon_2={1\over r}$, with $r$  the radius of the sphere.
   We will show that by restricting the overall sum of the masses in the fundamental representation,
the integral over $a$ localizes around some critical points where the full instanton partition function collapses  to a very simple form.
Exact formulae for the instanton partition function and for the expectation value of the circular Wilson/'t Hooft loops will then be derived. The results provide a direct  check of S duality. In particular we will explicitly check that Wilson and 't Hooft loops are 
exchanged under S duality and that the gauge theory partition function is S-duality invariant.  
\\
  We present also a qualitative study of the matrix model integral  in the case where some of the masses become large but keeping finite the overall sum. In this limit the integral can be computed with a saddle point approximation giving
  the main qualitative features of the localization for critical masses. The analysis follows closely   \cite{Passerini:2011fe,Russo:2012ay,Buchel:2013id,Russo:2013qaa} and shares with these studies   the leading  behavior of the Wilson loops in the limit of  a large number of colours.

This is the plan of the paper: In section 2 we  derive the partition function of
${\cal N}=2$  SYM on $\C^2$ with  fundamental matter  for some critical choices of the masses. In section 3 we treat the case of the 
gauge theory on $S^4$ and derive exact formulae for the partition function and the expectation value of
the Wilson and 't Hooft loops. In section 4 we describe the AGT dual of the gauge theory. The technical material
needed to  follow our computations is confined in the appendices. In particular,  appendix A collects the definitions and properties of 
the special functions used in the main text, appendix B the details of the four point function in Liouville theory
associated to the $SU(2)$ case, and appendix C is a detailed derivation of the main instanton partition function building blocks
for the gauge theory.

\section{The gauge theory on $\C^2$}

 We consider a four-dimensional  $U(N)$ gauge theory with $2N$ hypermultiplets, one half of these transforming in the
 fundamental representation and the other half in the anti-fundamental representation.
 \begin{center}
\includegraphics[width=0.50\textwidth]{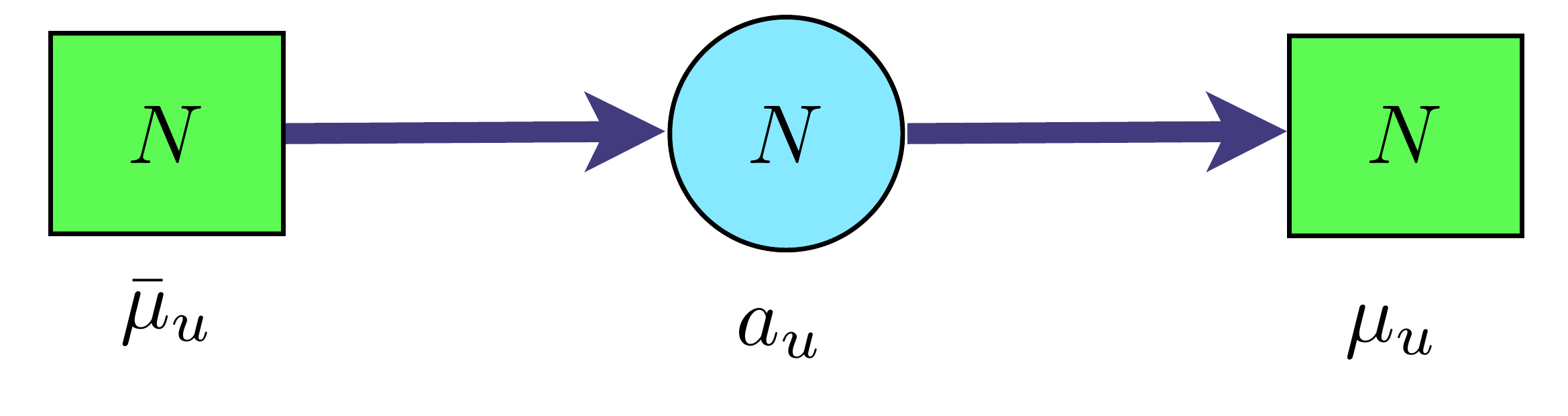}
%\caption{\label{tetrahedron}\small The tetrahedron in $\theta$-space.}
\end{center}
The  partition function of the gauge theory  is given by a product of the classical, one loop and instanton contributions
 \be
 Z=Z_{\rm class} \, Z_{\rm one-loop}\, Z_{\rm inst}
 \ee
 Denoting by $q=e^{2\pi i \tau}$ the gauge coupling and by $a=\{ a_u \}$ the vevs, the classical contribution reads
\be
Z_{\rm class}=q^{- {a \cdot a\over 2 \epsilon_1\epsilon_2}}
\label{zclass}
 \ee
  The instanton partition function  follows from the localization formula \cite{Nekrasov:2002qd,Flume:2002az,Bruzzo:2002xf}
  \be
Z_{\rm inst}
=\sum_{Y}   q^{|Y|} \,   \prod_{u,v=1}^N  { Z_{\emptyset,Y_v}(\bar \mu_u-a_v  ) \, Z_{Y_u,\emptyset}(a_u-\mu_v )\over
Z_{Y_u,Y_v}(a_u-a_v) }
 \label{zinst0}
\ee
 with $a_u,\mu_u,\bar\mu_u$ parametrizing the vev's and masses and the sum running over the array $Y=\{ Y_u \}$ of  $N$ Young tableaux specifying the positions of instantons around
 $a_u$.  We denoted by $|Y|$ the instanton number given by the total number of boxes in $Y$. Finally $Z_{Y_u,Y_v}$
 is given by \cite{Flume:2002az,Bruzzo:2002xf} (see appendix C for details)\footnote{
 We notice the  reflection symmetry
$ Z_{Y_u Y_v }(-\epsilon-x)=Z_{Y_v Y_u} (x)$ that implies that $Z_{\rm inst}$ is symmetric under the exchange
$\bar \mu_u  \leftrightarrow \mu_u-\epsilon$.}
  \bea
Z_{Y_u,Y_v}(x)  =&&     \prod_{ (i,j)\in Y_u}
      (x+\epsilon_1(i- k_{vj})-\epsilon_2 \, (j-1-\tilde k_{ui}))  \nn\\
 \times  &&  \prod_{ (i,j)\in Y_v}
      (x-\epsilon_1(i-1- k_{uj})+\epsilon_2 \, (j-\tilde k_{vi}))
        \label{zalbe}
\eea
with $(i,j)$ running over  rows and columns respectively of  either the $Y_u$ or $Y_v$ tableaux.
Here   $\{ k_{uj} \} $  and $ \{ \tilde k_{ui} \}$ are infinite and weakly decreasing sequences of positive integers
giving the length of the rows and columns respectively of the tableau $Y_u$\footnote{If $i$ ($j$) is greater than the number of rows (columns) in $Y_u$,
the value of $k_{uj}$ ($\tilde k_{ui}$) is zero.}.
 It is convenient to rewrite formula (\ref{zalbe})   in the infinite product form \cite{Billo:2013fi}
   \bea
Z_{Y_u,Y_v}(x)  &= &      \prod_{i,j=1}^{\infty }{}'\,
    {  x+\epsilon_1(i- k_{vj})-\epsilon_2 \, (j-1-\tilde k_{ui}) \over  x+\epsilon_1 i-\epsilon_2 \, (j-1)       }
  \label{zalbe2}
\eea
 where
 \be
\prod_{i,j=1}^\infty {}' \equiv   \lim_{L\to \infty}  \, (-L \epsilon_1)^{|Y_v|}  \,
(-L \epsilon_2)^{|Y_u|}    \prod_{i,j=1}^L \label{zalbe3}
 \ee
 One can check that the contributions
coming from the numerator and denominator in (\ref{zalbe2})   cancel against each other
except for a finite number of terms in the numerator, which reproduce (\ref{zalbe}),
 and in the denominator, which cancel the prefactors
in (\ref{zalbe3})\footnote{For example for $Y_u=\Yfund$, $Y_v=\emptyset$ one finds  $\lim_{L\to \infty}
 \, (-L \epsilon_2)
 \left({x+\epsilon\over x+\epsilon-L \epsilon_2}\right)=x+\epsilon$ in agreement with (\ref{zalbe}).}.
 Moreover, the prefactors in (\ref{zalbe3}) cancel in (\ref{zinst0}) between numerator and denominator and therefore the prime
 in the infinite product can be omitted. \\
 Interestingly the denominator in  (\ref{zalbe2}) does not depend on the shape of the Young tableaux
 and therefore this infinite product can be factored out. The one-loop contribution
is defined in such a way to cancel this $Y$-independent term.    Up to $a$-constant terms one can write
 \bea
Z_{\rm one-loop }
 &=&    \prod_{i,j=1}^\infty
    {  \prod_{u,v=1}^N ( \bar \mu_u -a_v+\epsilon_1 i-\epsilon_2 \, (j-1)   ) (    a_u -\mu_v +\epsilon_1i-\epsilon_2 \, (j-1)  )
    \over     \prod_{u<v}^N (a_{uv}+\epsilon_1 i-\epsilon_2 \, (j-1) ) (-a_{uv}+\epsilon_1 i-\epsilon_2 \, (j-1) )
     }\nn\\
 &=&   {  \prod_{u,v}  \Gamma_2(\bar \mu_u-a_v +\epsilon  ) \Gamma_2(a_u- \mu_v+\epsilon)
 \over   \prod_{u<v}      \Gamma_2(a_{uv} ) \Gamma_2(a_{uv} +\epsilon)   }  \label{zloop}
 \eea
 where in the second line we used (\ref{barnes2}) and (\ref{reflection}) to rewrite the infinite product in terms of the Barnes double Gamma
  function $\Gamma_2(x|\epsilon_1,\epsilon_2)$ (see appendix \ref{appendixA} for definitions and corresponding properties).
  Here and below  $\epsilon=\epsilon_1+\epsilon_2$.
  \\
 The prepotential of the gauge theory in the $\epsilon$-background is defined as \cite{Nekrasov:2002qd}
 \be
 {\cal F}=-\epsilon_1 \epsilon_2 \log Z
 \ee
    We notice that  the $Z_{\rm one-loop }$ given by (\ref{zloop}) has zeros in the moduli space of masses of the gauge theory.
   These zeros are located at the points where a flavor brane comes close to a gauge brane (for small $\epsilon_i$'s) and indicate that
  a fundamental matter is getting ``massless"\footnote{In presence of an $\epsilon$-background and vevs, the parameters $\mu,\bar \mu$   are related but not directly identified with the masses.}. In the following we will study the gauge theory in the nearby
  of these critical points.

\subsection{ Critical choices of masses}
\label{critmasses}

 The contribution of the fundamental and anti-fundamental matter to the instanton partition function
  can be written according to (\ref{zalbe}) as
 \bea
  Z_{\emptyset,Y_v}(\bar \mu_u-a_v) &=&            \prod_{(i,j)\in Y_v}
    (  \bar \mu_u-a_v -\epsilon_1( i-1) -\epsilon_2 \, (j-1) )\nn\\
 Z_{Y_u,\emptyset}(a_u-\mu_v ) &=&            \prod_{(i,j)\in Y_u}
    (  a_u-\mu_v +\epsilon_1i +\epsilon_2 \, j ) \label{zfund}
 \eea
   We notice that some eigenvalues in (\ref{zfund}) become zero for the critical choice of masses
  \bea
   \mu_u  &=& a_u+ p_u\, \epsilon_1+q_u\, \epsilon_2     \qquad p_u,q_u\in \Z_{\geq 1}
  \eea
  or
   \bea
  \bar \mu_u  &=& a_u + (p_u-1)\, \epsilon_1+(q_u-1)\, \epsilon_2     \qquad p_u,q_u\in \Z_{\geq 1}
  \eea
with $(p_u,q_u)$ some integers. Indeed for both choices,
   (\ref{zfund}) vanishes if the tableaux $Y_u$ contains the box $(i,j)=(p_u,q_u)$.
    In particular for   $(p_u,q_u)=(1,1)$ no  instanton contributions are allowed, for $(p_u,q_u)=(1,n)$ only those tableaux, $Y_u$,    with less than $n$ rows contribute and so on.

We will focus on the two simplest non trivial  choices
 \bea
{\bf I} && \mu_u=\mu^{\bf I}_u=a_u+ \epsilon         \nn\\
{\bf II} && \mu_u=\mu^{\bf II}_u=a_u+\epsilon+ \epsilon_2  \, \delta_{u,1}
\label{choices12}
 \eea
  with $\bar \mu_u$ arbitrary. In the case ${\bf I}$ one finds that all non-trivial Young tableaux give a vanishing contribution
leading to
  \be
  Z_{\rm inst}^{\bf I}=1   \label{zinst1}
  \ee
  On the other hand, for the choice ${\bf II}$ a non-trivial contribution  arises only from an array of tableaux with a single
  non empty tableau $Y_1$ made of a unique row and $Y_{u \neq 1}=\emptyset$.
For this simple configuration
 the instanton partition function can  be explicitly evaluated and resummed. Indeed, denoting by $k$ the length
 of $Y_1$, one finds
   \bea
   \label{ZFgen}
  Z^{\bf II}_{\rm inst}&=& \sum_{k=0}^\infty  q^k  \prod_{i=1}^k { \prod_{u=1}^N
     (   \bar \mu_u -a_1-\epsilon_1(i- 1)  )    (   a_1-\mu^{\bf II}_u +\epsilon_1 i +\epsilon_2 )  \over
   (    \epsilon_1 (i-k)+ \epsilon_2 )  (i \epsilon_1)  \prod_{u=2}^N    (   a_{1u}+\epsilon_1 i+ \epsilon_2 )  (   a_{u1}-\epsilon_1(i- 1)  ) }     \nn\\
     &=&
   \sum_{k=0}^\infty  q^k \, { \prod_{v=1}^N
         \left[ A_v \right]_k \over  k! \prod_{v=2}^{N} \left[ B_{v}\right]_k}  = {}_{N} F_{N-1}(^{\bf A}_{\bf B}\big| q)
      \eea
with $[x]_k=\prod_{p=0}^{k-1}(x+p ) $  , ${}_{N} F_{N-1}(^{\bf A}_{\bf B}\big| q) $ the generalized hypergeometric function\footnote{For the $U(1)$ case the generalized hypergeometric solution reduces to $(1-q)^{-{\bf A_1}}$.} and
 \bea
 \label{FAB}
 {\bf A}_v &=& { a_{1}  -\bar \mu_v  \over \epsilon_1  } =   { \mu_{1}  -\bar \mu_v -2\epsilon_2 \over \epsilon_1  }-1 \qquad~~~~~~~~~~~~~~ v=1,\ldots N \nn\\
  {\bf B}_{v}&=& {a_1-a_{v}+\epsilon_2\over \epsilon_1}+1={\mu_1-\mu_{v}\over \epsilon_1}+1
  \qquad~~~~~~~~~~~ v=2,\ldots N
\eea
Similar formulae can be found by replacing $\delta_{u,1}$ by $\delta_{u,j}$ in (\ref{choices12}), i.e.
\be
{\bf II} \qquad \mu^{(j)}_u=a_u+\epsilon+ \epsilon_2  \, \delta_{u,j}  \quad  \Rightarrow   \quad
Z^{\bf II}_{\rm inst,j}= {}_{N} F_{N-1}\left(^{{\bf A}^{(j)}}_{{\bf B}^{(j)}} \big| q \right) \label{ZFgen2}
\ee
with
\be
~~~A_k^{(j)}=1-B_j+A_k,~~~~~~~~~~~B_k^{(j)}= \left\{
\begin{array}{lc}
1-B_j+ B_k &~~~~ k \ne j\\
2-B_k&~~~~k=j\\
\end{array}\right.
\label{ABj}
\ee
and $B_1=1$. Formulae (\ref{ZFgen}) and (\ref{ZFgen2})  provide us with the simplest examples of ${\cal N}=2$ gauge theories  on $\C^2$
  where non-trivial multi-instanton corrections can be computed in an analytic form.  In section \ref{agtdual} we will see how
  this simplification can be understood from the point of view of the AGT dual Toda CFT where the instanton partition
  function is associated to correlators involving the insertion of degenerated fields. For example, the
  critical choice $(p_1,q_1)=(1,1)$, $(p_2,q_2)=(p,q)$ for  the $SU(2)$ gauge theory can be associated to the insertion of
  the so called $\phi_{(p,q)} $-degenerated field in Liouville theory.

\section{The gauge theory on $S^4$}

In this section we consider the gauge theory on $S^4$. The partition function in this case is  obtained by squaring  the $\Omega$-deformed gauge 
theory partition function $Z(a)$ on $\C^2$  and integrating it  over the vevs $a\in SU(N)$ of the scalar fields at infinity  \cite{Pestun:2007rz}. 
For a round sphere of radius $r$ one takes $\epsilon_1=\epsilon_2={1\over r}$, while  arbitrary $\epsilon_i$ represent the gauge theory 
on an ellipsoid \cite{Alday:2009aq}.
 
 In this section we derive exact formulae for the partition function of ${\cal N}=2$ gauge theories on ellipsoids with critical masses.   We will also compute  the expectation value of circular supersymmetric Wilson/'t Hooft loops relying on the localization formulae \cite{Pestun:2007rz,Alday:2009fs,Gomis:2011pf}.
 
\subsection{The partition function}

%%%%%%%%%%%%%%%%%%%%%%%%%%%%%%%%%%%%%%%%%%%%%%%%%%%%%%%
\begin{center}
\begin{figure}
\center
\includegraphics[width=0.8\textwidth]{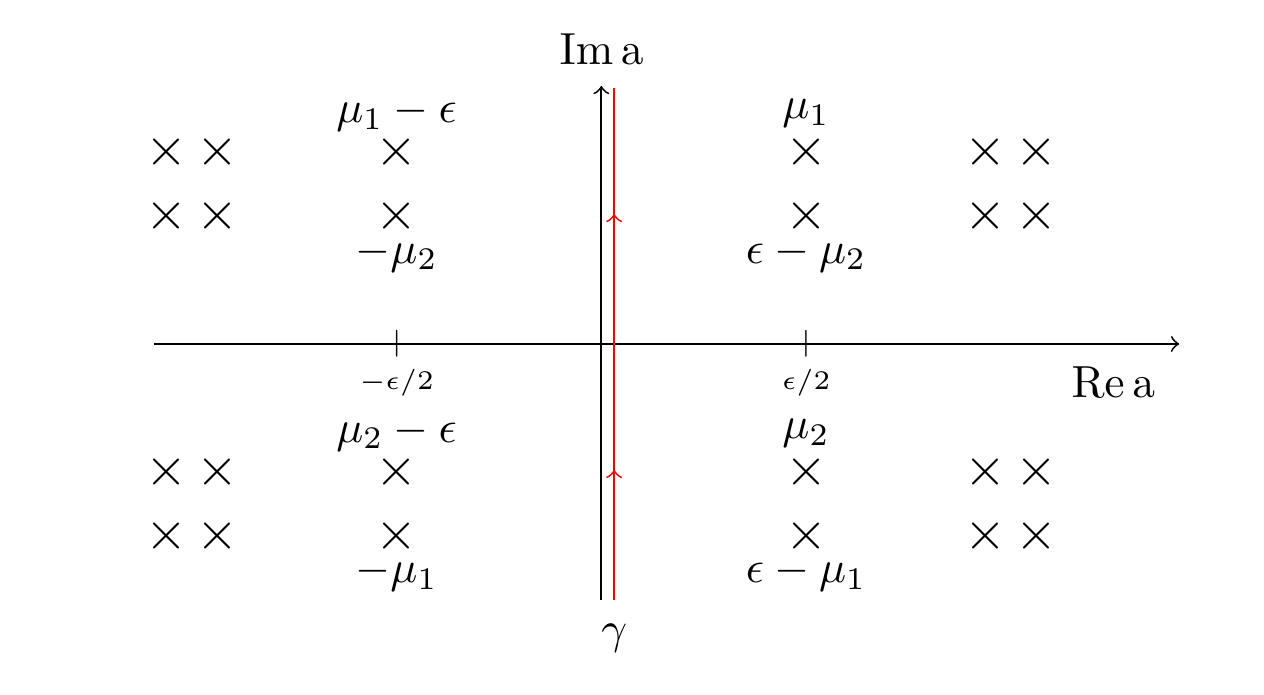}
\caption{ $\mu$-dependent poles  of $Z_{pert}(a)$ for the case of gauge group $SU(2)$. }
\label{fig1}
\end{figure}
\end{center}
%%%%%%%%%%%%%%%%%%%%%%%%%%%%%%%%%%%%%%%%%%%%%%%%%%%%%%%%%%%%%%%%%%%

 The partition function on  the sphere $S^4$ (or ellipsoid in the case $\epsilon_1\neq \epsilon_2$) is given by the integral \cite{Pestun:2007rz}
 \be
Z_{S^4}  =Z_{\rm flavor} \int_\gamma  da\,    | Z_{\rm class}Z_{\rm one-loop}Z_{\rm inst}  |^2 \label{zs4}
\ee
  with $da=\prod_{u=1}^{N-1} da_u$, $a_N=-\sum_{u=1}^{N-1} a_u$.  The integral runs over the lines  $a_u\in {\rm i} \R $.
 We take (consistently with the $\bar \mu \leftrightarrow \mu-\epsilon$ symmetry)  
    \be
   \bar \mu_u \in -\ft{\epsilon}{2}+ {\rm i} \R     \qquad  ,   \qquad \mu_u \in \ft{\epsilon}{2}+ {\rm i} \R  \label{range}
   \ee
  Notice that these conventions for the domain of $\bar \mu$'s  are different from that in \cite{Alday:2009aq}. This leads to a different look for  $Z_{\rm one-loop}$
   but  identical results for the truly physical quantity $|Z_{\rm one-loop}|^2$.
  
    The term $Z_{\rm flavor}$ is a normalization factor associated to the  $SU(N)^2\times U(1)^{2}$  flavor symmetry
    %\footnote{For a complete matching with the Toda conformal field theory it is necessary to add an extra normalisation factor $\Upsilon'(0)^{2N-2}$ in $Z_{\rm flavor}$.}
    \bea
 Z_{\rm flavor}=   \Upsilon(\kappa_0 )    \Upsilon( \kappa_1 )
 \prod_{u<v}  \Upsilon( \bar \mu_{uv}   )  \Upsilon( -\mu_{uv}   ) \label{zflavor}
 \eea
 with
 \be
 \kappa_0 = \sum_{u=1}^N (a_u-\bar \mu_u)   \qquad    \kappa_1= \sum_{u=1}^N (a_u-\mu_u+\epsilon)
 \ee
The function $\Upsilon(x)$ is defined in appendix A.
  Including or not this factor may be matter of taste since $Z_{\rm flavor}$  is independent of the $SU(N)$ gauge variables and it is therefore irrelevant to the dynamics of the non-abelian gauge theory. Still, the inclusion  of this term, as we will see,  guarantees the analyticity of the  partition function $Z_{S^4}$  over the
  $\mu,\bar \mu$  plane. In addition, the  gauge theory partition function defined in this way precisely matches, as we will see, the AGT dual 
  4-point correlator in  $A_{N-1}$ Toda field theory up to $\mu,\bar\mu$-independent constants.

 Together with the  classical and one loop
   contributions  (\ref{zclass})  and (\ref{zloop})  one finds
  \bea
 Z_{\rm pert}(a)&=& Z_{\rm flavor}|Z_{\rm class} Z_{\rm one-loop}|^2\nn\\
&    =&
  |q|^{-{a\cdot a\over \epsilon_1 \epsilon_2}}\,  { \Upsilon(\kappa_0) \Upsilon(\kappa_1)
 \prod_{u<v} \Upsilon(\bar \mu_{uv}) \Upsilon(a_{uv}  ) \Upsilon(a_{vu}) \Upsilon(\mu_{vu} )  \over
  \prod_{u,v=1}^N   \Upsilon(a_v-\bar \mu_u   ) \Upsilon(\mu_v -a_u) }   \label{zpert}
  \eea
%  The result can be rewritten in the compact and suggestive form
%  \be
%   Z_{\rm pert}  (a)   =  |q|^{2a\cdot a\over \epsilon_1 \epsilon_2}    \,
%   c( -\bar \mu, a,\kappa_0)c(-a, \mu,\kappa_1)
% \ee
% with
% \be
%  c(a,b,\kappa) =
% { \Upsilon(\kappa)
% \prod_{u<v} \Upsilon(a_{vu}) \Upsilon(b_{vu} )  \over
%  \prod_{u,v=1}^N   \Upsilon(a_u+b_v )  } \label{zpert2}
%  \ee
%
%
%

  We notice that the integrand in (\ref{zs4}) has an infinite number of poles in the $a$-plane.
  In Fig.\ref{fig1}, for the purpose of exemplification, we show the poles coming from
  the zeros of $\Upsilon(\mu_v -a_u)$  in the case of gauge group $SU(2)$. A similar sequence of poles comes from the zeros
  of  $\Upsilon(a_v-\bar \mu_u   )$.
   The integration in (\ref{zs4})
is along the path, given by the imaginary axis, which is marked in red in Fig.\ref{fig1}. We notice that for $\mu,\bar \mu$
in the range (\ref{range}) no poles fall along the contour $\gamma$   and moreover the number of poles on the two sides of $\gamma$ coincides.

The partition function $Z_{S^4}$, viewed as a function of the masses,
can be extended analytically to the whole
complex plane $\mu_u,\bar \mu_u \in \C$. In doing this the contour $\gamma$ in (\ref{zs4}) should be properly deformed
in such a way that the poles of the integrand  do not cross the integration path  $\gamma$.
Equivalently one can keep the contour always along the imaginary axis adding to the result of the integral  the sum of the residues over the poles  of the integrand crossing  $\gamma$. This procedure is well known from the CFT side \cite{Zamolodchikov:1995aa}  (see also \cite{Alday:2009fs}).

In the following, we  focus  on special choices of  the mass parameters of the gauge theory  for which the integral over $a$ in $Z_{S^4}$ can be 
analytically  computed. We consider two choices, that we refer as cases ${\bf I}$ and ${\bf II}$. They are closely related to the cases ${\bf I}$ and ${\bf II}$
  of the gauge theory on $\C^2$ considered in the last section. Indeed, as we will see, for the two choices the integral along $\gamma$ localizes  around 
  critical values  of $a_u$ satisfying the relations  (\ref{choices12})  for which the full instanton  sums $Z_{\rm inst}$  have been evaluated.
\\\\

\subsubsection{Critical masses: case {\bf I}}

\begin{center}
\begin{figure}
 \includegraphics[width=1\textwidth]{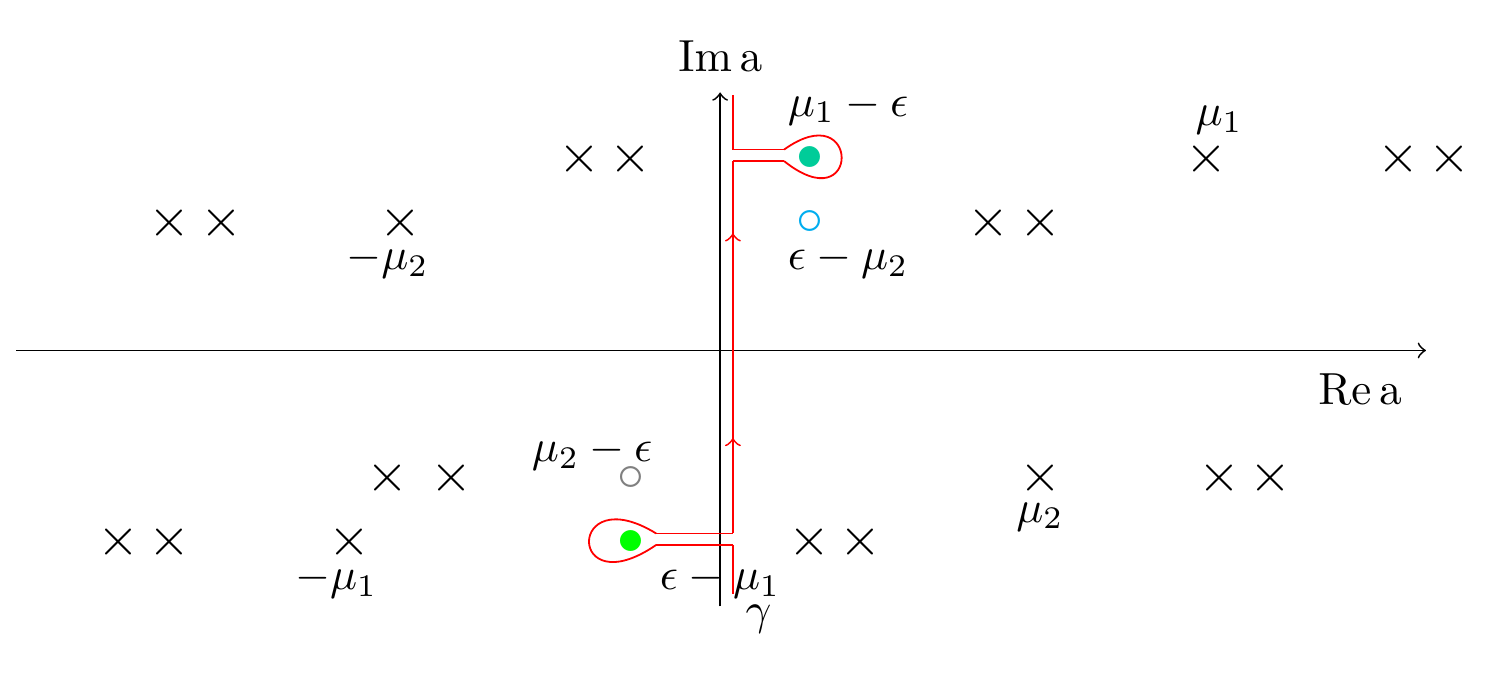}
 \caption{To obey to the condition (\ref{k1nepsilon}) the poles of Fig.\ref{fig1} must be shifted and the integration contour $\gamma$
needs to be deformed. In this figure we depict the situation for the gauge group $SU(2)$.}
\label{fig2}
\end{figure}
\end{center}

We consider first the co-dimension one slice of the moduli space defined by masses $\mu_u$ satisfying the relation   
\be
\kappa_1=\sum_{u=1}^N (\epsilon-\mu_u) = 0   \label{k1nepsilon}
\ee
 For this choice, $Z_{\rm one-loop} \sim \Upsilon(\kappa_1) = 0$ all the way along the $a$-plane except at the points where the denominator of
  $Z_{\rm one-loop}$ also vanishes.
   The integral is then given by the sum of the residues of those poles of $Z_{\rm one-loop}$ crossing $\gamma$ once we move from the
region of definition (\ref{range}). For the case ${\bf I}$  the  relevant poles and the contour are displayed in Fig.\ref{fig2}
for the gauge group $SU(2)$.
 This figure can be obtained from Fig.\ref{fig1} by applying a rigid shift to  $\mu_1$ and $\mu_2$ such that
 $\mu_1+\mu_2 = 2 \epsilon$. Under this shift the rows containing $\mu_{1,2}$ will be shifted towards
the right  while  those containing
 $-\mu_{1,2}$ will be shifted towards the left. In the process the poles denoted by a bullet, $\bullet$,
in the figure cross the  imaginary axis, ${\rm Im}\, a$, and  contribute to the residues. On the other hand,
the integrand in (\ref{zs4}) contains the terms $\Upsilon^{-1}(\mu_2-a_v)$  associated to the poles
labelled $\epsilon-\mu_2$ in the figure and denoted by $\color{cyan}\circ$, which come close to $\gamma$ from the other side and cancel 
the zero coming from  $\Upsilon(\kappa_1)$, leading to a finite result. A similar analysis can be performed for
the $SU(N)$ case. For simplicity,  we will take $\mu_u,\bar \mu_u$ from now on given by real numbers (with small imaginary parts) 
and therefore all the poles in the figures are located near the real line. 

Summarizing, for the gauge group $SU(N)$ the partition function gets  contributions only from the residue at the poles
 \be
  a_u  =a_u^{\rm crit}=\mu_{u} -\epsilon
  \ee
up to a permutation of the $\mu_u$'s.  The residue is given by

\be
Z^{\bf I}_{S^4}  = N\,N!  \,  {\rm Res}_{a=a^{\rm crit}} \left( Z_{\rm pert}(a)   \,|Z_{\rm inst}(a)|^2  \right)
=  |q|^{-{(\mu-\epsilon)^2\over \epsilon_1 \epsilon_2}}\,  { N\,N!\,\Upsilon(\kappa_0) \over \Upsilon'(0)^{N-1}}
 {\prod_{u<v} \Upsilon(\bar \mu_{uv})  \Upsilon(\mu_{vu} )  \over
  \prod_{u, v}   \Upsilon(\mu_v-\bar \mu_u -\epsilon  )   }
%   \nn\\
%&=& |q|^{2 (\mu-\epsilon)^2\over \epsilon_1 \epsilon_2}      \, {N! \over \Upsilon'(0)^{N-1} } \,
%   c(- \bar \mu, \mu-\epsilon,\kappa_0  )
 \label{zexact1}
\ee\\
  where we have used $Z_{\rm inst} (a_{\rm crit}^{\bf I} )=1$ according to
  (\ref{zinst1}). The $N!$ comes from the sum over the permutations of the $\mu_u$'s, while the extra $N$ counts their combinations in the
$N$-tuples which define each pole. Formula
  (\ref{zexact1}) is obvious from a CFT perspective since the critical choice ${\bf I}$ of masses corresponds to the insertion of
  an identity operator  leading to  a three (rather than four) point function that is clearly independent from the coordinates. On the other hand, the result is highly non-trivial from the gauge theory point of view, since it provides an exact formula for the gauge theory partition function on that slice of the moduli space defined by  (\ref{k1nepsilon}).

%%%%%%%%%%%%%%%%%%%%%%%%%%%%%%%%%%%%%%%%%%%%%%%%%%%%%%%%%%%%%%%%%%%%
%%%%%%%%%%%%%%%%%%%%%%%%%%%%%%%%%%%%%%%%%%%%%%%%%%%%%%

%  \subsection{The perturbative expansion}
%
% Let us consider first the limit $\tau_2 \to \infty$ for generic masses. For simplicity we specify to $SU(2)$ gauge theory
% on a round sphere, i.e. $b=1$. The perturbative part of the gauge theory partition function
%  becomes
%  \bea
%  Z_{S^4, {\rm pert}}&=& Z_{\rm flavor} \int da  \,e^{-4 \pi a^2 \tau_2} \,  {
%  \Upsilon( 2 a  ) \Upsilon(-2 a)   \over
%   \prod_{u}  \Upsilon(\bar \mu_u-a )\Upsilon(\bar \mu_u+a ) \Upsilon(a-\mu_u )\Upsilon(-a-\mu_u ) } \nn\\
%   &=& { Z_{\rm flavor} \Upsilon'(0)^2 \over
%   \prod_{u}  \Upsilon(\bar \mu_u )^2   \Upsilon(-\mu_u )^2  } \int da (-4 a^2) \left[1+ c_2 \, a^2 +\ldots\right]
%  \eea
%  with
%  \be
%  c_2=\sum_{u=1}^2 \left( {\Upsilon'(\bar \mu)^2\over \Upsilon(\bar \mu)^2}+{\Upsilon'(-\mu)^2\over \Upsilon(-\mu)^2}
%  -  {\Upsilon''(\bar \mu)\over \Upsilon'(\bar \mu)}-{\Upsilon''(-\mu)\over \Upsilon(-\mu) } -{\Upsilon''(0)^2\over \Upsilon'(0)^2}
%+  \ft43 {\Upsilon'''(0)\over \Upsilon'(0)}   \right)
%  \ee
%
%
%  \be
%  \Upsilon(x)=G(x) G(2-x)
%  \ee
%   and $G(x)$ the Barnes function, satisfying  $G(x+1)=\Gamma(x) G(x)$ and

  \subsubsection{Critical masses: case II}

  The next simplest (but non-trivial) case corresponds to the choice
  \be
\kappa_1=\sum_{u=1}^N (\epsilon-\mu_u)=- \epsilon_2
\ee
 Once again,  $Z_{\rm one-loop} \sim \Upsilon(\kappa_1)= 0$ almost everywhere except at the  poles
  in (\ref{zpert}). The situation is very similar to the previous case. The relevant  poles can now be written as
  \be
  a_u  =a_{u,j}^{{\rm crit}}=\mu_{u} -\epsilon-\delta_{uj} \, {\epsilon_2}
  \label{aucrit}
  \ee
  up to  permutations of the $\mu_u$'s.
  The partition function becomes
\be
Z^{\bf II}_{S^4} = \sum_{j=1}^N   c_j  |Z_j(q)|^2   \label{zj}
\ee
with
\be
\label{ratioCC}
Z_{j}=Z_{\rm inst}(a^{(j)}_{\rm crit}) =   {}_NF_{N-1}\left(^{{\bf A}^{(j)} }_
{{\bf B}^{(j)}} \big| q\right)
\ee
written in terms of the ${\bf A}^{(j)} , {\bf B}^{(j)} $   defined in (\ref{ABj}) and (\ref{FAB}), and
\bea
c_j &=&N! \,{\rm Res}_{a= a^{(j)}_{{\rm crit}} }  Z_{\rm flavor}(a)| Z_{\rm class}  Z_{\rm one-loop}(a)|^2 \label{cjz}\\
  &=&  N! \,
  |q|^{-{a^{(j)}_{\rm crit} \cdot a^{(j)}_{\rm crit}\over \epsilon_1 \epsilon_2}}\,  {\, \Upsilon(\kappa_0)\prod_{u<v} \Upsilon( \bar \mu_{uv})
\Upsilon( \mu_{vu} ) \over   \Upsilon'(0)^{N-1} \prod_{u, v}   \Upsilon(   \mu_v -\bar \mu_u-\epsilon -\epsilon_2 \delta_{vj} )        }
 \prod_{u \neq j}  {  \Upsilon( \mu_{uj}  - \kappa_1  )  \over
 \Upsilon(
 \mu_{uj}  )      }\nn
   \eea
     Alternatively one can write
     \be
Z^{\bf II}_{S^4} = c_1 \, \sum_{j=1}^N   r_j  |Z_j(q)|^2   \label{zj2}
\ee
with
\be
r_j = {c_j\over c_1}= |q^{1-B_j} |^2 \prod_{u=1}^{N}\frac{\gamma(B_u)\,\gamma(B_{j}-B_u)}{\gamma(A_u)\,\gamma( B_{j}-A_u)}
\label{ratios}
\ee
and  the $\gamma(x)$ function defined in (\ref{gammadef}).  Written in this form, one can check that the $Z_{S^4}$  defined by  (\ref{zj2}) with $r_j$ given by (\ref{ratios}) is single valued on the whole $q$-complex plane. This fact is obvious around the point $q=0$ given that $Z_j$  picks up a phase when going around the origin. The single valuedness around $q= \infty$ follows by sending $q \to q^{-1}$ and using the trigonometric relation (\ref{fnn}) and the form of $r_j$ to show that $Z_{S^4}$ is again given by a sum of squares around $q= \infty$. This provides a highly non-trivial consistency check of the results for $Z_{S^4}$. In the section \ref{agtdual} we will present a further test of this result by matching the gauge theory partition
function with the AGT dual correlator in the $A_{N-1}$Toda Field Theory.

\subsection{Wilson loops}

%  The   Wilson loop
%   \be
% W  =\  e^{i \int_C A_M dx^M}
% \ee
% with   $A_M$ a six-dimensional vector comprising the gauge $A_\mu$ and the
%  scalar field $\varphi=\varphi_1+{\rm i} \varphi_2$  is said to be supersymmetric if it is annihilated by the equivariant BRS charge $Q$. A simple choice of supersymmetric Wilson loop is given by
%  \be
%{\rm i}  \int_C A_M dx^M ={  \pi {\rm i}  \varphi_\epsilon \over \epsilon_1}+{\rm h.c.} ={\pi {\rm i}   \over \epsilon_1}(\varphi+A_\mu \Omega^\mu{}_\nu x^\nu)    + {\rm h.c.}
%  \ee
%  with $\Omega^\mu{}_\nu$ an antisymmetric  matrix with $\Omega_{12}=\epsilon_1$, $\Omega_{34}=\epsilon_2$.
%
The expectation value of a supersymmetric circular Wilson loop (at the equator in the first plane) is given by the localization formula
\cite{Pestun:2007rz,Alday:2009fs}
      \be
\langle W \rangle  =\left\langle e^{i \int_C A_M dx^M}  \right\rangle  =
\frac{1}{Z_{S^4} } \int_\gamma  da\,    Z_{\rm pert}(a) | Z_{\rm inst}(a) |^2    {\rm Tr}_{\bf R} \, e^{2\pi i a\over \epsilon_1 }
\label{wloop}\ee
  Plugging (\ref{zexact1}) and (\ref{zj2}) into (\ref{wloop}) one finds for the two choices of critical masses and for ${\bf R}$ being the 
fundamental representation
 \bea
 \langle W \rangle^{\bf I}
 &=&  \sum_{u=1}^N e^{2\pi i {(\mu_u-\epsilon )\over \epsilon_1} }
    \nn\\
\langle W \rangle^{\bf II}
 &=&  { \sum_{j,u=1}^N e^{2\pi i {(\mu_u-\epsilon- \epsilon_2\delta_{uj} ) \over \epsilon_1} }  \, r_j |  Z_j   |^2  \over  \sum_{j=1}^N  r_j\,|  Z_j    |^2 }
    \label{wilsoni}
\eea
 Formulae (\ref{wilsoni}) are exact and show that $\ln \langle W \rangle  $ grows linearly with the perimeter
 $2\pi/ \epsilon_1$ of the Wilson loop. Interestingly on a round sphere both results reduce to the classical contribution
 $\langle W \rangle=  \sum_{u=1}^N e^{2\pi i \mu_u}$.

\subsection{'t Hooft  loops}

The results in the last sections are exact and therefore can be used to study the strong coupling behavior of the partition
function and Wilson loops in ${\cal N}=2$ gauge theories with critical masses. Indeed, using standard transformation properties of the  
generalized hypergeometric functions the results can be rewritten as an expansion around $q= 1$ rather than $q= 0$. Here 
we illustrate this analysis for the $SU(2)$ case. \\ 
Let us consider first the partition function. Using (\ref{idf21}) and (\ref{euler}) (and dropping the subscript 2 for $B$) one can write
\small{
\bea
Z_1 &=&\frac{\pi \Gamma(B)}{\sin(\pi(B-A_1-A_2))}\left(\frac{\tilde{Z}_1}{\Gamma(B-A_1)\Gamma(B-A_2)\Gamma(A_1+A_2-B+1)}  \right.\nn\\
&& \left.- \frac{ (1-q)^{B-A_1-A_2} \,   \tilde{Z}_2 }{\Gamma(A_1)\Gamma(A_2)\Gamma(B-A_1-A_2+1)}\right)\nn\\
Z_2 &=&\frac{\pi \Gamma(2-B) \,q^{B-1} }{\sin(\pi(B-A_1-A_2))}\left(
\frac{\tilde{Z }_1 }{\Gamma(1-A_1)\Gamma(1-A_2)\Gamma(A_1+A_2-B+1)}
\right.\nn\\
&& \left.- \frac{ (1-q)^{B-A_1-A_2} \,  \tilde{Z}_2 }{\Gamma(1+A_1-B)\Gamma(1+A_2-B)\Gamma(B-A_1-A_2+1)}\right)
\label{GGtilde}
\eea
}
with
\bea
 Z_1 &=& {}_2F_{1}\left(^{~~A_1 ,A_2 }_
{~~B} \big| q\right)   \qquad ~~~~~~~~~~~    Z_2= {}_2F_{1}\left(^{1-B+A_1,1-B+ A_2 }_
{~~~~2-B} \big| q\right) \nn\\
\tilde Z_1 &=& {}_2F_{1}\left(^{~~A_1 ,A_2 }_
{A_1+A_2-B+1} \big| 1-q\right)   \qquad      \tilde Z_2= {}_2F_{1}\left(^{B-A_1,B- A_2 }_
{1+B-A_1-A_2} \big| 1-q\right) 
\eea
 We notice that the quantities with the tilde are obtained from those without via the replacements\footnote{The replacement $q \to 1-q$ corresponds to send $\{\infty,1,q,0\}$ into $\{\infty, 0, q, 1\}$, while $B\to A_1+A_2-B+1$ amounts to exchange  $\alpha_2 \leftrightarrow \alpha_4$ on the CFT side, according to the notation in \ref{CFTgaugedictionary}.}
\be
q \to 1-q   \qquad B \to A_1+A_2-B+1 \label{replacement}
\ee 
Plugging (\ref{GGtilde}) into (\ref{zj2}) one finds
\be
Z_{S^4}= c_1  |Z_1|^2+c_2 |Z_2|^2  =\tilde c_1  |\tilde Z_1|^2+\tilde c_2 |\tilde Z_2|^2 \label{zstrong}
\ee
with 
\footnotesize
\be
\tilde c_1=  {  \gamma(B) \gamma(1-B+A_1) \over \gamma(B-A_2) \gamma(1-B+A_1+A_2)}  \,c_1 \qquad 
\tilde c_2= {  |1-q|^{2(B-A_1-A_2)}\, \gamma(B) \gamma(A_1+A_2-B) \over \gamma(A_1)  \gamma(A_2)\gamma(B-A_2)\gamma(1-B+A_2)}  \,c_1 
\qquad 
\ee 
\normalsize
The equivalent descriptions in the left and right hand sides of (\ref{zstrong}) provide the expansions of the gauge theory
partition function in the weak ($q\approx 0$) and strong ($q\approx 1$) coupling regimes.  The natural variables in the two regimes are
related by the map (\ref{replacement}).

Finally, we consider the expectation value of  a circular Wilson loop  in the fundamental representation (spin $\ft12$) of $SU(2)$. 
Following  \cite{Pestun:2007rz,Alday:2009fs} the evaluation of the circular Wilson loop boils down
to  insert the operator $2 \cos \left(\frac{2\pi a }{\epsilon_1}\right)$ in the gauge theory partition function. After localization around 
$a=a^{(j)}_{\rm crit}=\{ \ft12 (B-1)\epsilon_1\mp \ft12 \epsilon_2 \}$ this results into
\be
 \langle W \rangle = {1\over Z_{S^4}}\,\left[  -2\, c_1   \cos \left( \pi B-\ft{\pi \epsilon_2}{\epsilon_1} \right)\,    |Z_1|^2-2\,c_2 \cos \left( \pi B+\ft{\pi \epsilon_2}{\epsilon_1} \right) \, |Z_2|^2 \right] \label{ww}
\ee
Using (\ref{GGtilde}) to rewrite (\ref{ww}) in terms of the $\tilde Z_i$ one finds for the expectation value of the 't Hooft loop
\be
  \langle H_{1/2} \rangle = {c_{ij} \tilde Z_i \overline{\tilde{Z}}_j   \over Z_{S^4}}
\ee
with
{\footnotesize
\bea
c_{11}&=&{2\tilde c_1 \left(    \cos \left( \pi B+\ft{\pi \epsilon_2}{\epsilon_1} \right)\,   
 \sin{\pi A_1} \sin{\pi A_2}   -   \cos \left( \pi B-\ft{\pi \epsilon_2}{\epsilon_1}\right) \sin{\pi (B-A_1)} \sin{\pi (B-A_2)} \right)  \over \sin{\pi B}  \sin{\pi (B-A_1 -A_2)}\,   
}\label{cij}\\
 c_{22}&=&{2\tilde c_2 \left( \cos \left( \pi B-\ft{\pi \epsilon_2}{\epsilon_1} \right)\,   
 \sin{\pi A_1} \sin{\pi A_2} - \cos \left( \pi B+\ft{\pi \epsilon_2}{\epsilon_1} \right)\,   
 \sin{\pi (B-A_1)} \sin{\pi (B-A_2)}  \right)   \over \sin{\pi B}  \sin{\pi (B-A_1 -A_2)}}\nn\\
c_{21} &=&{ (1-q)^{B-A_1-A_2}\,2\pi^3\, \tilde c_1\, \left(\cos \left( \pi B-\ft{\pi \epsilon_2}{\epsilon_1}\right)-\cos \left( \pi B+\ft{\pi \epsilon_2}{\epsilon_1} \right)  \right) \over \sin \pi B \sin^2 \pi(B-A_1-A_2)\,   \Gamma(B-A_1-A_2) \Gamma(1+B-A_1-A_2) \prod_{i=1}^2 \Gamma(A_i)   \Gamma(1-B+A_i) } \nn\\
c_{12} &=&{(1-q)^{A_1+A_2-B}\, 2\pi^3\, \tilde c_2\, \left( \cos \left( \pi B-\ft{\pi \epsilon_2}{\epsilon_1} \right)-\cos \left( \pi B+\ft{\pi \epsilon_2}{\epsilon_1} \right)\right) \over \sin \pi B \sin^2 \pi(B-A_1-A_2)\, \Gamma(A_1+A_2-B) \Gamma(1-B+A_1+A_2)  \prod_{i=1}^2   \Gamma(1-A_i)   \Gamma(B-A_i)   }\nn
\eea
}
The quantities $c_{ij}$ are related to  the $H_{\pm}$, $H_0$ obtained in \cite{Alday:2009fs} from a CFT analysis 
(see \cite{Gomis:2011pf} for direct 
computations of 't Hooft loops in the gauge theory side). In appendix \ref{sthooft} we provide a detailed
comparison of the results. The matching provides a strong consistency test of our results here and it is an explicit check of S duality among Wilson and 't Hooft loops.%further supports the AGT 
%description of 't Hooft loops developed in \cite{Alday:2009fs} by a direct computation in the gauge theory. 
 \subsubsection{A saddle point analysis}

The qualitative behavior of the results we found for the expectation value of the Wilson loop in ${\cal N}=2$ gauge theories with critical masses can be confirmed by considering the limit in which some of the masses are large keeping the overall sum small.
 This limit can be treated with a simple saddle point analysis along the lines of  the large $N$ analysis in \cite{Passerini:2011fe,Russo:2012ay,Buchel:2013id,Russo:2013qaa}.
As we will see, this rather crude approximation already captures the localization of the $a$-integral and the
qualitative behaviour of our result for gauge theories with critical masses.
  For simplicity, we focus on the SU(2) case.  We write
\be
Z_{\rm pert}(a)=e^{-S_{\rm eff}}
\ee
  with
 \be
S_{\rm eff} \approx-\ln  \Upsilon(2a) \Upsilon(-2a)  +\sum_{u=1}^2  \ln\Upsilon(a-\bar \mu_u   )
\Upsilon(\mu_u -a)\Upsilon(-a-\bar \mu_u   ) \Upsilon(\mu_u +a)+\frac{a^2}{\epsilon_1\epsilon_2}\ln |q|
\ee
For $\epsilon_1=\epsilon_2=1$, and $\mu_1=-\mu_2=\mu$ with $\mu >> \bar\mu_u$, the  expansion (\ref{upsilonlargex})
leads to
\bea
S_{\rm eff} &=&-2\,a^2\ln (2a)^2+  (\mu -a)^2\ln(\mu -a)^2
+(\mu +a)^2\ln(\mu +a)^2 +\ldots  \label{seff}
 \eea
    The leading contribution to the integral of $|Z_{\rm pert} Z_{\rm inst}|^2$ comes from the $a_{\rm crit}$
    extremizing $S_{\rm eff}$, i.e.  from the solutions  of $S_{\rm eff}'(a_{\rm crit})=0$. One finds $a_{\rm crit}=\pm \mu$.
  Plugging into   (\ref{wloop}) one finds for the expectation value of a Wilson loop in the fundamental representation
  \be
    \langle W \rangle  \approx     2 \cos 2\pi \mu
   \ee
 in agreement with (\ref{wilsoni}) for a round sphere $\epsilon_1=\epsilon_2=1$.

 \subsection{The AGT dual}
\label{agtdual}
\begin{figure}[t]
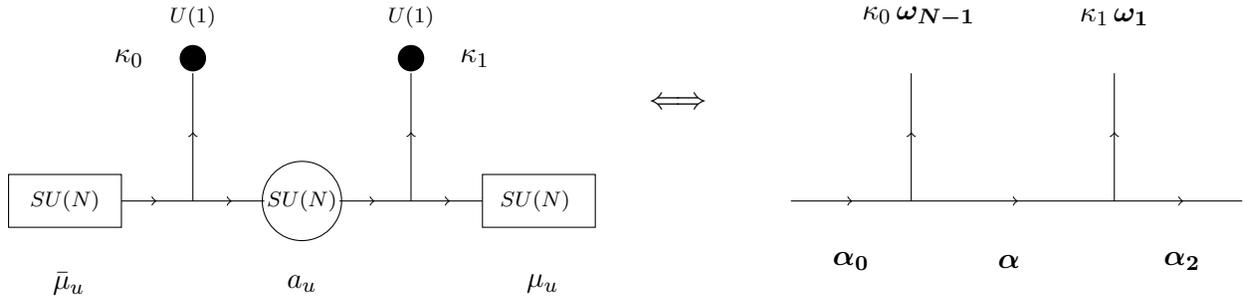

\begin{pgfpicture}{0cm}{0cm}{15cm}{6cm}
%% diagramma sinistra
\pgfsetendarrow{\pgfarrowto}

\pgfcircle[stroke]{\pgfpoint{4cm}{2.3cm}}{15pt}
\pgfputat{\pgfxy(4,2.3)}{\pgfbox[center,center]{\scriptsize{$SU(N)$}}}
\pgfputat{\pgfxy(4,1.2)}{\pgfbox[center,center]{\small{$a_u$}}}
{\color{black}\pgfcircle[fill]{\pgfpoint{2.55cm}{4.2cm}}{5pt}}
\pgfputat{\pgfxy(2.55,4.75)}{\pgfbox[center,center]{\scriptsize{$U(1)$}}}
\pgfputat{\pgfxy(1.7,4.2)}{\pgfbox[center,center]{\small{$\kappa_0$}}}
{\color{black}\pgfcircle[fill]{\pgfpoint{5.45cm}{4.2cm}}{5pt}}
\pgfputat{\pgfxy(5.45,4.75)}{\pgfbox[center,center]{\scriptsize{$U(1)$}}}
\pgfputat{\pgfxy(6.3,4.2)}{\pgfbox[center,center]{\small{$\kappa_1$}}}
{\color{black}\pgfrect[stroke]{\pgfpoint{0.1cm}{1.95cm}}{\pgfpoint{1.5cm}{20pt}}}
{\color{black}\pgfputat{\pgfxy(0.85,2.3)}{\pgfbox[center,center]{\scriptsize{$SU(N)$}}}}
\pgfputat{\pgfxy(0.9,1.2)}{\pgfbox[center,center]{\small{$\bar{\mu}_u$}}}
%\pgfrect[stroke]{\pgfpoint{1cm}{4cm}}{\pgfpoint{1.5cm}{20pt}}
{\color{black}\pgfrect[stroke]{\pgfpoint{6.4cm}{1.95cm}}{\pgfpoint{1.5cm}{20pt}}}
{\color{black}\pgfputat{\pgfxy(7.1,2.3)}{\pgfbox[center,center]{\scriptsize{$SU(N)$}}}}
\pgfputat{\pgfxy(7.2,1.2)}{\pgfbox[center,center]{\small{${\mu}_u$}}}
%\pgfrect[stroke]{\pgfpoint{5.5cm}{4cm}}{\pgfpoint{1.5cm}{20pt}}
\pgfline{\pgfxy(1.6,2.3)}{\pgfxy(2.05,2.3)}
\pgfline{\pgfxy(2.05,2.3)}{\pgfxy(3.05,2.3)}
\pgfline{\pgfxy(4.5,2.3)}{\pgfxy(4.95,2.3)}
\pgfline{\pgfxy(4.95,2.3)}{\pgfxy(5.95,2.3)}
\pgfline{\pgfxy(2.55,2.3)}{\pgfxy(2.55,3.2)}
\pgfline{\pgfxy(5.45,2.3)}{\pgfxy(5.45,3.2)}

\pgfclearendarrow

\pgfline{\pgfxy(3.05,2.3)}{\pgfxy(3.5,2.3)}
\pgfline{\pgfxy(5.95,2.3)}{\pgfxy(6.4,2.3)}
\pgfline{\pgfxy(2.55,3.2)}{\pgfxy(2.55,4)}
\pgfline{\pgfxy(5.45,3.2)}{\pgfxy(5.45,4)}

\pgfputat{\pgfxy(9,3.6)}{\pgfbox[center,center]{$\Longleftrightarrow$}}

%% diagramma destra

\pgfsetendarrow{\pgfarrowto}

\pgfline{\pgfxy(10.5,2.3)}{\pgfxy(11.3,2.3)}
\pgfline{\pgfxy(11.3,2.3)}{\pgfxy(13.5,2.3)}
\pgfline{\pgfxy(13.5,2.3)}{\pgfxy(15.7,2.3)}
\pgfline{\pgfxy(12.1,2.3)}{\pgfxy(12.1,3.2)}
\pgfputat{\pgfxy(13.4,1.5)}{\pgfbox[center,center]{\small{$\boldsymbol{\alpha} $}}}
\pgfputat{\pgfxy(12.2,4.75)}{\pgfbox[center,center]{\small{$\kappa_0\,\boldsymbol{\omega_{N-1}}$}}}
\pgfputat{\pgfxy(11.3,1.5)}{\pgfbox[center,center]{\small{$\boldsymbol{\alpha_0} $}}}
\pgfline{\pgfxy(14.8,2.3)}{\pgfxy(14.8,3.2)}
\pgfputat{\pgfxy(14.8,4.75)}{\pgfbox[center,center]{\small{$\kappa_1\,\boldsymbol{\omega_1}$}}}
\pgfputat{\pgfxy(15.7,1.5)}{\pgfbox[center,center]{\small{$\boldsymbol{\alpha_2} $}}}

\pgfclearendarrow

\pgfline{\pgfxy(15.7,2.3)}{\pgfxy(16.5,2.3)}
\pgfline{\pgfxy(12.1,3.2)}{\pgfxy(12.1,4)}
\pgfline{\pgfxy(14.8,3.2)}{\pgfxy(14.8,4)}
\end{pgfpicture}
\caption{On the left the quiver diagram for the conformal $SU(N)$ gauge theory: in evidence the  $SU(N)^2\times U(1)^2$ flavor factors. On the right the diagram of the conformal block for the dual Toda field theory.}
\label{figAGT}
\end{figure}

 In this section we describe the AGT dual of the gauge theory partition function with critical masses. According to
 the AGT dictionary  \cite{Alday:2009fs,Wyllard:2009hg}, the partition function of an $SU(N)$ gauge theory with
 $N$ fundamental and $N$ anti-fundamental hypermultiplets is mapped to a four point function on the sphere in
 $A_{N-1}$  Toda field theory with two $U(1)$ and two $SU(N)$ punctures, see Fig. \ref{figAGT}.
 The choice of critical masses corresponds to the insertion of a degenerated field at one of the $U(1)$ punctures.  The corresponding correlator was computed in \cite{Fateev:2005gs,Fateev:2007ab}. Here we collect some background material
 on the Toda Field theory and the results for the relevant correlator showing the matching with the gauge theory answer.

\subsubsection{Toda field theory}

Let us briefly recall few facts concerning the Toda field theory. The action is given by
\bea
{\cal A}=\int d^2 x \left(\frac{1}{8\pi} (\partial \varphi)^2+m
\sum_{i=1}^{N-1}e^{b(e_i,\varphi)}\right)
\eea
where $\varphi$ is a $N$ dimensional vector equipped with the usual Euclidean scalar product, whose components sum
to zero and  $e_i$ are the simple roots of the algebra $A_{N-1}$. Explicitly
\be
(e_i)_k=\delta_{i,k}-\delta_{i+1,k}
  \ee
  with $k=1,2,\ldots N$. The fundamental weights $\omega_i$, by definition,
constitute the dual basis: $(e_i,\omega_j)=\delta_{i,j}$. Explicitly
\bea
(\omega_i)_{k}=
\left\{
\begin{array}{cll}
1-\frac{i}{N},& k \leq i\\
-\frac{i}{N},& k>i
\end{array}
\right.
\eea
We also need the set of weights of the fundamental representation
\bea
h_k=\omega_{1}-e_1-\cdots -e_{k-1}
\eea
for $k>1$ and $h_1=\omega_1$. In particular, for an arbitrary vector $a$ one finds
\be
(h_k,a)=a_k-\frac{1}{N}\sum_{k=1}^N a_{k}    \qquad    (h_k,h_l)=\delta_{k,l}-{1\over N}
\ee
Let us define ${\bf Q}=Q \rho$, where $Q=b+1/b$ and
\bea
\rho=\frac{1}{2}\sum_{k=1}^N (N+1-2 k)h_k
\eea
is the Weyl vector (half sum of all positive roots). The central charge of
the Virasoro algebra is
\bea
c=N-1+12 ({\bf Q},{\bf Q})=(N-1)(1+N(N+1)Q^2)
\eea
The primary fields, $V_{\boldsymbol\alpha}=\exp
(\boldsymbol\alpha,\phi)$, in the Toda theory
have conformal dimension
\bea
\Delta_{\boldsymbol\alpha}=\frac{(\boldsymbol\alpha,2{\bf Q}-
\boldsymbol\alpha)}{2}= {({\bf Q}-P,{\bf Q}+P)\over 2}    \nonumber
\eea
with $P$ the momentum defined by
\be
P=\boldsymbol\alpha-{\bf Q}
\ee
 Conventionally the states are parameterized in terms of $P$
and the fields via $\boldsymbol\alpha$.  It is important that this
parametrization is modulo Weyl reflections, which simply permute the components
of the momenta.
We also notice that ${\boldsymbol \alpha}$ can always be chosen such that $\sum_u \alpha_u=0$ and then
\be
(\boldsymbol\alpha-{\bf Q}, h_u)=(\boldsymbol\alpha-{\bf Q} )_u
\ee
%The eigenvalue of the zero mode
%of first non-trivial W-current $W^{(3)}(z)$ (in $SU(N)$ Toda there are
%holomorphic currents of spin $s$, $W^{(s)}$, with $s=3,\ldots,N$) in
%terms of the momentum $P$ is given by
%\bea
%w=\frac{i}{3}\sqrt{\frac{2N}{(N-2)(4+N(N+2)Q^2)}}\,\,\sum_{i=1}^NP_i^3
%\eea
The W-Ward identities (there are holomorphic currents of spin $s$, $W^{(s)}$, with $s=3,\ldots,N$) show that a two point function
$\langle V_1(z_1) V_2(z_2) \rangle$ is nonzero only if $P_1=-P_2$. An ingoing state should then be accompanied by the reflection $P \rightarrow -P$
or equivalently
$\boldsymbol\alpha \rightarrow {2\bf Q}-\boldsymbol\alpha$.

\subsubsection{The four point conformal block}

  The instanton partition function of the gauge theory on $\mathbb{C}^2$ can be  related (see \ref{CFTgaugedictionary} for details) 
  to the chiral four point function conformal block\footnote{   We notice that $V_{\kappa_1\omega_{1}}  \sim
 V_{(N\epsilon-\kappa_1)\omega_{N-1}}$. Indeed the momenta
$\kappa_1\omega_{1}-{\bf Q}$ and $(N\epsilon-\kappa_1)\omega_{N-1} -{\bf Q}$
 are related by a cyclic permutation of components.    } 
 %multiplied by $U(1)$ factor\footnote{The chiral conformal block is the expression (\ref{GV}) divided by $\langle V_{2{\bf Q}-{\boldsymbol\alpha}_0}(\infty))V_{\kappa_0\omega_{N-1}}(1)V_{{\boldsymbol\alpha}}(0)\rangle=1$ and $\langle V_{2{\bf Q}-{\boldsymbol\alpha}}(\infty))V_{\kappa_1\omega_{1}}(z)V_{{\boldsymbol\alpha_2}}(0)\rangle=z^{\Delta_{{\boldsymbol\alpha}}-\Delta_{\kappa_1\omega_{1}}-\Delta_{{\boldsymbol\alpha_2}}}=z^{ (h_1,\boldsymbol{\alpha}_2) b}$. For $N=2$ the term $(1-z)^{\frac{ \kappa_0}{ N}b} $ is canceled by multiplying by the $U(1)$ factor: see \ref{CFTgaugedictionary} for details.}
\be
 \langle V_{2{\bf Q}-{\boldsymbol\alpha}_0}(\infty)V_{\kappa_0\omega_{N-1}}(1)
V_{\kappa_1 \omega_{1}}(z)V_{{\boldsymbol\alpha}_2}(0)\rangle  = z^{ (h_1,\boldsymbol{\alpha}_2) b}(1-z)^{\frac{ \kappa_0}{ N}b}   G(z)
\label{GV}
\ee
 with   $\kappa_1$ given by
\be
\kappa_{1}=(1-p) {1\over b}+(1-q) b
\ee
for some $p,q\in \Z_{\geq 0}$. These operators   are associated to degenerated
fields in the Toda field theory. A correlator involving the insertion of a degenerated field  satisfies a
differential equation of order $pq$. In particular the case $p=q=1$ corresponds to  the identity operator
and the associated conformal block is trivial
\bea
{\bf I} &&  \kappa_1=0  \qquad G(z)=1
\eea
The first interesting case occurs at $(p,q)=(1,2)$ associated to the so called $\phi_{12}$ field.
The function $G(z)$ in this case satisfies the Pochhammer differential equation (see \cite{Fateev:2005gs,Fateev:2007ab})
\be
\left[ z\prod_{i=1}^N\left(z\frac{\partial}{\partial z}+A_i\right)- z
\frac{\partial}{\partial z}\prod_{i=2}^{N}\left(z\frac{\partial}{\partial z}+B_i-1\right)\right] G(z)=0
\ee
with
\bea
&&A_j=\frac{\kappa_0}{ N} b+\frac{(1-N)}{ N}b^2+(h_1,\boldsymbol{\alpha_2}-{\bf Q})b-(h_j,\boldsymbol{\alpha_0}-{\bf Q})b \quad j=1,\ldots N   \nn\\
&&B_j=1+  (h_1-h_{j},\boldsymbol{\alpha_2}-{\bf Q})b~~~~~~\quad j=2,\ldots N   \label{abs}
\eea
The $N$ independent solutions  can be written as
\bea
{\bf II} &&  \kappa_1= -b: \qquad G_{j}(z)=z^{1-B_j}{}_NF_{N-1}\left(^{{\bf  A}^{(j)} }_
{{\bf  B}^{(j)}    }\big| z\right)\label{cftii}
\eea
with $j=1,\ldots,N$, $B_1=1$ and  ${\bf  A}^{(j)},{\bf  B}^{(j)}$ given by (\ref{ABj}).

\subsubsection{The four point correlator}

%Alternatively,  a dual basis of solutions with diagonal monodromy around $z=\infty$ can be written as
%\bea
%H_{k}(z)=z^{-A_k}{}_NF_{N-1}\left(^{1+A_k-B_1,\ldots ,1+A_k-B_{N-1} }_
%{1+A_k-A_1,\ldots ,1+A_k -A_{k-1} , 1+A_k-A_{k+1},\ldots ,1+A_k-A_{N}}\big| {1\over z}\right)
%\eea
The full correlator can be written as
\be
{\cal G}(z,\bar z)=\sum_{j=1}^N  C_j \left|  z^{(h_1,\boldsymbol{\alpha}_2) b}(1-z)^{\frac{ \kappa_0}{ N} b}  G_j(z) \right|^2 \label{cgg}
%=\sum_{k=1}^N  \tilde{C}_k |H_k(z)|^2
\ee
with $C_j$  determined by the requirement that ${\cal G}(z,\bar z)$ is single valued  over the $z$-plane. Alternatively
the coefficients $C_j$ can be built out of the three point vertices
\be
C_j= {1 \over (2\pi i)^{N-1}} \oint_{\boldsymbol{\alpha^{(j)}_{{\rm crit}}} } d^{N-1}\alpha\,  C(2{\bf Q}-\boldsymbol{\alpha_0},\boldsymbol{\alpha},\kappa_0 w_{N-1} ) \, C(2{\bf Q}-\boldsymbol{\alpha},\boldsymbol{\alpha_2},\kappa_1 w_{1} )
\ee
with\footnote{For simplicity we work in units  where $  \pi \mu \gamma(b^2) b^{2-2b^2} =1$.}
\be
C({2\bf Q}-\boldsymbol{\alpha},\boldsymbol{\beta},\kappa w_{N-1} )= { \Upsilon(b)^{N-1} \Upsilon(\kappa)
 \prod_{u<v} \Upsilon( (\boldsymbol{\alpha}-{\bf Q},h_u-h_v ))
\Upsilon( ({\bf Q}-\boldsymbol{\beta},h_u-h_v )) \over    \prod_{u,v} \Upsilon( {\kappa\over N}+
 ({\bf Q}-\boldsymbol{\alpha},h_u)+   (\boldsymbol{\beta}-{\bf Q},h_v)  ) }\label{c3}
\ee
and the same formula for the three point function involving the insertion of $V_{\kappa w_1}$ with  $h_{u} \to -h_{u}$,
$h_{v} \to -h_{v}$ in the denominator of (\ref{c3}).
The contour integral picks up the residue at the pole
\be
\boldsymbol{\alpha^{(j)}_{{\rm crit}}}=\boldsymbol{\alpha_{2}}+\kappa_1 \, h_j
\ee
with $\kappa_1=- b$.
One finds
 {\small \bea
C_j &=& {\rm Res}_{\boldsymbol{\alpha}= \boldsymbol{\alpha^{(j)}_{{\rm crit}}} }   C({2\bf Q}-\boldsymbol{\alpha_0},\boldsymbol{\alpha},\kappa_0 w_{N-1} )C({2\bf Q}-\boldsymbol{\alpha},\boldsymbol{\alpha_2},\kappa_1 w_{1} )\label{cjs}\\
&= &{\Upsilon(b)^{2N-2}\over \Upsilon'(0)^{N-1}   }   { \Upsilon(\kappa_0)  \prod_{u<v} \Upsilon( (\boldsymbol{\alpha_0}-{\bf Q})_{uv})
\Upsilon( ({\bf Q}-\boldsymbol{\alpha_2})_{uv} )  \prod_{u \neq v}  \Upsilon( (\boldsymbol{\alpha_2}-{\bf Q} )_{uv} +\kappa_1 \delta_{uj}- \kappa_1 \delta_{vj}) \over
   \prod_{u, v}   \Upsilon( {(\kappa_0-\kappa_1)\over N}+
 ({\bf Q}-\boldsymbol{\alpha_0},h_u)+   (\boldsymbol{\alpha_2}-{\bf Q} ,h_v) +\kappa_1 \delta_{jv} ) \prod_{u\ne v}\Upsilon(
 (\boldsymbol{\alpha_2}-{\bf Q})_{uv}  +\kappa_1\delta_{uj} )       }\nn\\
 &=& {\Upsilon(b)^{2N-2}\over \Upsilon'(0)^{N-1}   }
 { \Upsilon(\kappa_0)\prod_{u<v} \Upsilon( (\boldsymbol{\alpha_0}-{\bf Q})_{uv})
\Upsilon( ({\bf Q}-\boldsymbol{\alpha_2})_{uv} ) \over   \prod_{u, v}   \Upsilon( {(\kappa_0-\kappa_1)\over N}+
 ({\bf Q}-\boldsymbol{\alpha_0},h_u)+   (\boldsymbol{\alpha_2}-{\bf Q} ,h_v) +\kappa_1 \delta_{jv} )        }
 \prod_{u \neq j}  {  \Upsilon( (\boldsymbol{\alpha_2}-{\bf Q} )_{uj}  - \kappa_1  )  \over
 \Upsilon(
 (\boldsymbol{\alpha_2} -{\bf Q})_{uj}  )      } \nn
   \eea}
   Using the dictionary below (\ref{todagauge}) one can see that the result (\ref{cgg}) with $C_j$ given by (\ref{cjs})
   perfectly matches the gauge theory partition function up to an irrelevant constant. In particular, the ratios
$R_j={C_j\over C_1}$ are given again by (\ref{ratios}) with $A_j,B_j$ now given by  (\ref{abs}): $R_j=r_j |q^{B_j-1}|^2$. This ensures the single
valuedness of $G(z,\bar{z})$ and provides us with a highly non-trivial consistence check of the results for $C_j$.

\subsubsection{The Toda/gauge theory dictionary}

 CFT and gauge theory variables are related via the dictionary
   \bea
(P_m)_u &=& (\boldsymbol{\alpha_m}-{\bf Q})_u = a_{m ,u}-{1\over N} \sum_{u=1}^N a_{m,u}   \qquad a_{m,u}=(a_{0,u},a_{1,u},a_{2,u})=(\bar \mu_u,a_u,\mu_u)    \nn\\
 \kappa_{0} &=&  \sum_{u} (a_u-\bar \mu_u)\qquad \kappa_{1} =  \sum_{u} (a_u-\mu_u+\epsilon)  \qquad
 G_j(q)=   q^{ \mu_j-\mu_1   \over  \epsilon_1   } \,   Z_j(q)     \label{todagauge}
 \eea
 with $\epsilon_1={1\over b}$, $\epsilon_2= b$.
 Plugging   (\ref{todagauge}) into (\ref{cftii}) one finds perfect agreement
between the Toda conformal blocks $G_j(q)$  and the instanton partition functions $Z_{j}$
for the critical choices of masses  ${\bf II}$ (\ref{ZFgen2}). 
On the other hand the residues $C_j$ in the Toda field theory  reproduce the one-loop contributions in the gauge theory
on $S^4$.
Case ${\bf I}$ corresponds to the insertion of an identity operator and therefore the conformal block is
trivial consistently with the fact that $Z_{\rm inst}^I=1$  in the gauge theory side. 
\section{Summary of results}

In this paper we  have derived exact formulae for the  partition function  of a special class of ${\cal N}=2$ gauge theories on $\R^4$ and $S^4$  with fundamental and anti-fundamental matter.  On $\R^4$ we have showed that choosing the  difference between
the masses and the vev's to be a combination of the $\epsilon_\ell$ with integer coefficients, only a very restrictive type of  instanton configurations (Young tableaux shape) can contribute to the
gauge theory  partition function. The critical choices of masses are in correspondence with the choices of degenerated field
insertions in the AGT dual Liouville or Toda field Theory. The simplest non-trivial cases correspond to either no instantons, or
 instantons associated to Young tableaux made of a single row. In the latter case the full instanton partition function
 can be evaluated and written in terms of generalized hypergeometric functions.

 On $S^4$, a critical mass corresponds to discrete choices of the overall mass associated to the center $U(1)$  of the $U(N)$ flavor group acting on the fundamental matter. For these choices the integral $\int da |Z(a)|^2$ localizes around a finite set of
 critical points where it can be explicitly evaluated as a sum of residues of the integrand. The gauge theory computation mimics
 that of a very well known CFT correlator in Toda field theory involving the insertion of a degenerated field.
 Exact formulae for the expectation values of circular supersymmetric Wilson and 't Hooft loops  in this class of gauge theories are also
 derived.

\vspace{1cm}

\centerline{\large\bf Acknowledgments}

\vspace{0.5cm}

We thank M. Billo, M.L. Frau, A. Lerda, M. Matone and Y. Stanev for useful discussions.
This work is partially supported by the
ERC Advanced Grant n.226455 ``Superfields" and by the Italian MIUR-PRIN contract 2009-KHZKRX. \\
The research of R.P. was partially supported by Volkswagen foundation of Germany, by
 the grant of Armenian State Council of Science 13-1C278 and by the Armenian-Russian grant ``Common projects in Fundamental Scientific Research"-2013.\\
%20075ATT78.
 The research of D.R.P. is also supported by the Padova University Project CPDA119349.
% and by the MIUR-PRIN contract 2009-KHZKRX.

\vspace{0.5cm}

\begin{appendix}

\section{The Barnes double gamma function}

\label{appendixA}
The Barnes double gamma function can be defined by the integral\footnote{In the following and in the main text we used the shorthand notation $\Gamma_2(x)$ when it is not necessary to specify its dependence on the $\epsilon_i$. The
    related function $\gamma_{\epsilon_1,\epsilon_2}(x)=\Gamma_2(x+\epsilon)$ is often used in the literature.}
\be
\log\Gamma_2(x|\epsilon_1,\epsilon_2)  = {d\over ds}\left(  {\Lambda^s\over \Gamma(s) } \int_0^\infty {dt\over t}
{t^s\, e^{-x t} \over
(1-e^{-\epsilon_1 t} ) (1-e^{-\epsilon_2 t} )  }\right)\Big|_{s=0}  \label{ge1e2}
\ee
Using the representation of the logarithm
\be
\log {x\over \Lambda}=-{d\over ds}\left(  {\Lambda^s\over \Gamma(s) } \int_0^\infty {dt\over t}  t^s\, e^{-x t}  \right)\Big|_{s=0}
\ee
one can think of $\Gamma_2(x)$  as  a regularization of the infinite product (for $\epsilon_1,\epsilon_2>0$)
\be
\Gamma_2(x) =\prod_{i,j=0}^\infty   \left( {\Lambda \over x+i \epsilon_1+j\epsilon_2}\right)
\ee
   with poles in the non-negative integers $(i,j)$. Using the same manipulations one finds (for $\epsilon_1>0$, $\epsilon_2<0$)
   \bea
   &&  \prod_{i,j=1}^\infty
   \left( {
   x+i \epsilon_1-(j-1) \epsilon_2}\over  \Lambda \right)=\Gamma_2(x+\epsilon)
   %\nn\\
     % &&  \prod_{i,j=1}^\infty
   %\left( {
  %- x+i \epsilon_1-(j-1) \epsilon_2}\over  \Lambda \right)=\Gamma_2(x)
   %\nn\\
%   &&  \prod_{i,j=1}^\infty
%   \left( {
%   x-i \epsilon_1+(j-1) \epsilon_2}\over  \Lambda \right)=\Gamma_2(x)
 \label{barnes2}
   \eea
The following identity holds
\be
\Gamma_2(x+\epsilon_1)\, \Gamma_2(x+\epsilon_2)=x\, \Gamma_2(x)\, \Gamma_2(x+\epsilon_1+\epsilon_2)
\ee
For large ${\rm Re} \, x$ the expansion of (\ref{ge1e2}) can be written as
\bea
\log \Gamma_2(x) &=&  c_0 \, x^2\left( -\ft14 \log \left( {x\over \Lambda}\right)^2+\ft34\right) +c_1 \, x \left(  \ft12 \log \left( {x\over \Lambda}\right)^2 -1\right)\nn\\
&&-\ft{c_2}{4}
\log\left( {x\over \Lambda}\right)^2 +\sum_{n=3}^\infty  {c_n x^{2-n} \over n(n-1)(n-2) }\label{g2exp}
\eea
with $c_n$ defined by
\be
{1\over (1-e^{-\epsilon_1 t} ) (1-e^{-\epsilon_2 t} ) }=\sum_{n=0}^\infty  {c_n\over n!} t^{n-2}
\ee
i.e.
\be
c_0={1\over \epsilon_1 \epsilon_2}  \qquad   c_1={\epsilon_1+\epsilon_2\over  2 \epsilon_1 \epsilon_2}
\qquad c_2={\epsilon_1^2+3 \epsilon_1 \epsilon_2+\epsilon_2^2 \over 6\epsilon_1 \epsilon_2}
\ee
For ${\rm Re} \, x<0$ we defined $\Gamma_2(x)$ via the  (\ref{g2exp}),  so we have the simple reflection property
\be
\Gamma_2(-x|\epsilon_1,\epsilon_2)=\Gamma_2(x|-\epsilon_1,-\epsilon_2)=\Gamma_2(x+\epsilon|\epsilon_1,\epsilon_2)
\label{reflection}
\ee
where the last equality follows from (\ref{ge1e2}).

\subsection*{The case $\epsilon_1={1\over b}$, $\epsilon_2=b$}

We will mainly consider the case
\be
\epsilon_1={1\over b} \qquad   \epsilon_2=b    \qquad    Q=\epsilon=\epsilon_1+\epsilon_2={1\over b}+b
\ee
 and define
\be
\Upsilon(x)={1\over \Gamma_2(x|\ft{1}{b},b) \Gamma_2(Q-x|\ft{1}{b},b)  }
\ee
The function $\Upsilon(x)$ is an entire function with zeros at
\be
\Upsilon \left( -m b-\ft{n}{b} \right)=\Upsilon \left( (m+1) b+\ft{n+1}{b} \right)=0 \qquad {\rm for} \qquad m,n\in \Z_{\geq 0}
\ee
which satisfies the properties
\bea
\label{relgamma}
\Upsilon(x)&=&\Upsilon(Q-x)   \qquad ~~~~~~~~~~~~~~~~\Upsilon(\ft{Q}{2})=1\nn\\
\Upsilon(x+b) &=& \gamma(b\,x) b^{1-2\, b\,x} \, \Upsilon(x) \qquad ~\Upsilon(x+\ft{1}{b}) = \gamma(\ft{x}{b}) b^{ {2\,x\over b} -1} \, \Upsilon(x)\nn\\
\eea
with
\be
\label{gammadef}
\gamma(x)={\Gamma(x)\over \Gamma(1-x) }
\ee
The function $\Upsilon(x)$ admits the integral representation
\be
\log \Upsilon(x)=\int_0^\infty \frac{dt}{t}\left[ \left({Q\over 2}-x \right)^2 e^{-t} -{ \sinh^2 \left({Q\over 2}-x \right)\ft{t}{2}  \over  \sinh \ft{bt}{2}
 \sinh \ft{t}{2b} } \right]
\ee
The case $b=1$ is particularly simple. The double Gamma function reduces to the G-Barnes function $G(x)$
\bea
G(x) ={1\over \Gamma_2(x|1,1) }       \qquad    \Upsilon_{b=1}(x)=G(x) G(2-x)
\eea
 satisfying $G(x+1)=\Gamma(x) G(x)$, $G(1)=1$. The G-function can be expanded  around $x\approx 0$
 \bea
 \log G(x+1)=\ft{x}{2} ( \log 2\pi-1)-(1+\gamma) \ft{x^2}{2}+\sum_{n=3}^\infty (-)^{n-1}\zeta(n-1) {x^n\over n}
 \eea
or for large $x$
 \bea
 \log G(x+1)=x^2 ( \ft14 \ln x^2-\ft34)+\ft{x}{2}\ln 2\pi-\ft{1}{24} \ln x^2+\zeta'(-1)+\sum_{n=1}^\infty  {B_{2n+2}\over 4 n(n+1) x^{2n} }  \nn
 \eea
leading to
\bea
 \log  \Upsilon(x+1|1,1)&=& \log G(1+x)+\log G(1-x) \nn\\
 &=& x^2 ( \ft12 \ln x^2-\ft32) -\ft{1}{12} \ln x^2+2\zeta'(-1)+\sum_{n=1}^\infty  {B_{2n+2}\over 2 n(n+1) x^{2n} }
\label{upsilonlargex} \eea
for large $x$ and
\bea
 \log \Upsilon(x+1|1,1)= -\sum_{n=2}^\infty \zeta(2n-1) {x^{2n}\over 2n}-(\gamma+1)x^2
 \eea
 for $x$ small. In the previous formulae $\zeta(n)$ is the Riemann zeta function, the $B_{n}$ are the Bernoulli numbers, while $\gamma$ is the Euler-Mascheroni constant.

\subsection{Hypergeometric identities}

The large $z$ expansion of the hypergeometric functions can be found from the Taylor expansion around $z=0$ via the identity 
\be
  {}_{N} F_{N-1}(^{\bf A}_{\bf B}\big| z)  \prod_{i=1}^N \frac{\Gamma(A_i) }{\Gamma(B_i)}
= \sum_{k=1}^N (-z)^{-A_k} \Gamma(A_k)  {\prod_{i\neq k}^N \Gamma(A_i-A_k) \over
\prod_{i=2}^{N} \Gamma(B_i-A_k)}     {}_N F_{N-1} \left(^{1+A_k-{\bf \hat B} }_{ 1+A_k-{\bf \hat A}^{(k)} } \Big|
 \ft{1}{z}\right)\label{fnn}
 \ee
 with ${\bf \hat A}^{(k)}=\{A_{u\ne k}\}_{u=1,\ldots,N}$ and ${\bf \hat B}=\{1,B_2,\ldots,B_N\}$.
 A similar relation allows to write the expansion around $z=1$. Explicitly, for $N=2$ one finds
 \bea
&&\frac{\sin \left(\pi(B-A_1-A_2)\right)}{\pi \Gamma \left(B \right )} \,\,  {}_{2} F_{1}\left({}^{A_1,A_2}_{~~B}\Big|z\right) \label{idf21}\\
&&=\frac{1}{\Gamma \left(B-A_1\right) \Gamma \left(B-A_2\right) 
\Gamma \left(A_1+A_2-B+1 \right )}\,\,
{}_{2} F_{1}\left(   {}^{A_1,A_2}_{A_1+A_2-B+1}\Big|1-z\right)\nonumber\\
&&-\frac{(1-z)^{{B-A_1-A_2}}}{\Gamma \left(A_1\right)
\Gamma \left(A_2\right)\Gamma \left(B-A_1-A_2+1 \right )}
 \,\,   {}_{2} F_{1}\left( {}^{B-A_1,B-A_2}_{B-A_1-A_2+1}\Big|1-z\right)\nn
\eea
We will also use in the main text  the Euler's transformation
\be
{}_{2} F_{1}\left({}^{A_1,A_2}_{~~B}\Big|z\right)=(1-z)^{B-A_1-A_2} {}_{2} F_{1}\left({}^{B-A_1,B-A_2}_{~~B}\Big|z\right) \label{euler}
\ee

\section{The correlator in Liouville theory}
\label{sliouville}

In this appendix we  derive the four point function in the dual 2d Liouville CFT involving the insertion of a null state.  We consider a two dimension conformal field theory with  central charge
\be
c=1+6 Q^2 =1+6\left(b+{1\over b}\right)^2
\ee
stress energy tensor $T$ and chiral fields $\phi_h$. The operator product expansions (OPE) read
\bea
T(w) \phi_h(z) &\approx &{ h \phi_h \over (w-z)^2} +   { \partial_z \phi_h(z) \over (w-z)}+\ldots\nn\\
T(w) T(z) &\approx &{  c \over (z-w)^4} + { 2 T(z) \over (w-z)^2} +   { \partial_z T(z) \over (w-z)}+\ldots\label{ope}
\eea
We define the Virasoro operators
\be
L_n^z\, {\cal O} (z) =\oint_{\gamma_z} {dw\over 2\pi i} \, (w-z)^{n+1}  \,T(w) \, {\cal O}(z)    \label{to}
\ee
with $\gamma_z$ a contour around $z$. Virasoro generators satisfy the algebra
\be
[L_n,L_m]=(n-m)L_{n+m} +{c\over 12} (n^3-n)\delta_{n+m,0}
\ee
It is convenient to parametrize the dimension of a primary field as
\be
h(\alpha)=\alpha (Q-\alpha)
\ee
and denote the corresponding field as $\phi_{\alpha}$. It should be noted that
this parametrization is not one to one, since the fields $\phi_\alpha$ and $\phi_{Q-\alpha}$
are identified up to a numerical multiplier, called reflection amplitude.\\
A null state ${\cal O}_{\rm null}$ is defined by the condition
\be
L_n^z {\cal O}_{\rm null}(z)=0   \qquad n>0
\ee
The general null state is labeled by two integers and can be written as
\be
{\cal O}_{mn}={\bf L}_{mn} \phi_{mn}
\ee
 with ${\bf L}_{mn}$ being some polynomial of the Virasoro generators
 at the total level $N=n\, m$ acting on a primary field $\phi_{mn}$ of conformal
dimension
\be
h_{mn}=\alpha_{mn}(Q-\alpha_{mn})
  \ee
  with two equally admissible choices for the parameters $\alpha_{mn}$
  \be
     \alpha_{mn}\equiv  {1\over 2b} (1-m) +{b\over 2}(1-n)
\label{degalpha1}
   \ee
   or
\be
     \alpha_{mn}\equiv  {1\over 2b} (1+m) +{b\over 2}(1+n)
\label{degalpha2}
   \ee
For instance
\bea
{\cal O}_{12}&=&\left( L_{-1}^2+b^2  L_{-2} \right) \phi_{12}   \qquad h_{12}= -\ft12-\ft{3 b^2}{4}\\
{\cal O}_{13}&=&\left( L_{-1}^3+4  b^2 L_{-1} \, L_{-2} +(4 b^4-2b^2) L_{-3}\right) \phi_{13}   \qquad h_{13}= -1- 2b^2\nn
\eea
and so on.

\subsection{The degenerated conformal blocks}

We are interested in four point functions involving the insertion of a $\phi_{nm}$ primary state. We denote the four-point function by
\be
F (z_i)=\langle \phi_{1}(z_1)\phi_{2}(z_2) \phi_3(z_3) \phi_4(z_4) \rangle
\ee
 Conformal invariance strongly restricts the form of $F$.
 Using the OPE (\ref{ope}) one expects
\be
\langle T(w) \phi_{1}(z_1)\phi_{2}(z_2) \phi_3(z_3) \phi_4(z_4) \rangle  =\sum_{i=1}^4 \left(  {h_i F\over (w-z_i)^2 } +   {F_i\over (w-z_i) }   \right) \label{tg}
\ee
with $F_i=\partial_{z_i} F$. On the other hand at large $w$, the correlator (\ref{tg})
is expected to fall as $w^{-4}$ since the vacuum is annihilated by
$L_0$, and $L_{-1}$,  i.e. $\langle 0| T(w)=\sum_{n\in \Z} \langle 0| L_n w^{n-2}\sim w^{-4}$. Equating to zero the coefficients of $w^{-n}$ with $n=1,2,3$ one finds
\bea
\sum_{i=1}^4 F_i &=&0 \nn\\
\sum_{i=1}^4 (z_i F_i+h_i F)&=& 0\nn\\
\sum_{i=1}^4 (z_i^2 F_i+2 z_i h_i F )&=& 0   \label{ggg}
\eea
These equations imply in particular that $F$ is a function only of the harmonic ratio $z={z_{12} z_{34} \over z_{13} z_{24}}$.\vspace{0.07cm}
Moreover equations (\ref{ggg}) can be used  to expressed $F_1,F_2,F_4$ in terms of $F(z)$ and $F'(z)$. In particular
setting
\be
z_1=\infty \qquad z_2=1 \qquad z_3=z \qquad z_4=0
\ee
one finds
\be
F_1=0 \qquad F_2=-z\,F' +(2\,h_1-\delta)F   \qquad F_4=(z-1) F' +(\delta-2 h_1) F
\ee
with $\delta=\sum_i h_i$.  We are interested in correlators  of the type
\be
F(z)=\langle   \phi_{\alpha_1}(\infty) \phi_{\alpha_2}(1) \phi_{\alpha_3}(z) \phi_{\alpha_4}(0) \rangle
\ee
 with
 \be
 \alpha_3=-\ft{b}{2}
 \ee
such that $\phi_{\alpha_3}\equiv \phi_{(12)}$ is a degenerated  field at the second
level\footnote{We have chosen here to exploit the expression (\ref{degalpha1})
for the parameter $\alpha_{1,2}$. As we will see later, it is this choice which
corresponds to the case in which only tableaux with one row contribute to the gauge theory, as discussed in section \ref{critmasses}.}.
 Using the fact that the null state ${\cal O}_{12}$ is orthogonal to all the other states in the theory, one can write
\be
0=\langle   \phi_{\alpha_1}(\infty) \phi_{\alpha_2}(1){\cal O}_{12}(z) \phi_{\alpha_4}(0) \rangle
 ={\bf L}_{12} F=\left( L_{-1}^2+b^2\, L_{-2} \right) F(z) \label{diffg}
\ee
where
\bea
L_{-1}^n\, F &=&\partial_{z}^n\, F \nn\\
L_{-2} \,F  &=&  \sum_{i\in \{1,2,4\}}  \left[  {h_i F\over (z_3-z_i)^2 } +   {F_i\over (z_3-z_i) }   \right]
%\nn\\
%L_{-3} \,G  &=& - \sum_{i=2}^4  \left[ {2h_i G\over (z_1-z_i)^3 } +   {G_i\over (z_1-z_i)^2 }   \right]
\eea
 The forms of  $L_{-1,-2}$  follow from (\ref{to}) after using the OPE (\ref{ope}).
 Writing
 \be
 F(z)=z^{\alpha_4 b}\, (1-z)^{\alpha_2  b} f(z)
 \ee
 the differential equation (\ref{diffg}) reduces to the hypergeometric equation
\bea
z(1-z)f''(z)+(B-(A_1+A_2+1)z)f'(z)-A_1 \,A_2 \, f(z)=0
\label{2F1difeq}
\eea
with
\bea
A_1 &=&-\frac{b^2}{2}+(\alpha_2+\alpha_4-\alpha_1)b,
\,\, A_2=-1-\frac{3 b^2}{2}+(\alpha_1+\alpha_2+\alpha_4)b,\nonumber\\
B&=& 2\alpha_4 b-b^2 \label{bfab}
\eea
The two independent solutions of (\ref{diffg})\footnote{The two solutions $z^{-\alpha_4 b}\, (1-z)^{-\alpha_2  b} F^{(\pm)}(z)$ correspond to $G_{j}(z)$ in  (\ref{cftii}) with $j=1,2$.}
\bea
\label{12CFTblock1}
F^{(+)}(z) &=& z^{\alpha_4  b}\, (1-z)^{\alpha_2 b}   {}_2F_1(^{\bf A}_{ B}\big| z) \label{12CFTblock2} \\
F^{(-)}(z)&=&z^{1+ b^2-\alpha_4 b}\, (1-z)^{\alpha_2 b} \,{}_2F_1\left(^{1-B+{\bf A} }_{2-B}\Big| z\right)\nonumber
\eea
correspond to the exchange in the $s$-channel of the field with dimension specified by
the parameters
\be
\alpha_{\pm} =\alpha_4\mp \frac{b}{2}
\ee
 (these are the only primary fields which result in the fusion of the fields $\phi_{1,2}(z)$ and $\phi_{\alpha_4}(0)$ at $z=0$).

\subsection{The physical correlator}

The physical correlation function can be built out of the conformal blocks $F^{(\pm)}$ and the
three point function $C(\alpha_1,\alpha_2,\alpha_3) $ by the so standard gluing algorithm.
For the Liouville theory the three point function is given by
\bea
\label{Calphai}
&& C(\alpha_1,\alpha_2,\alpha_3) =\left( \pi \mu \gamma(b^2) b^{2-2b^2}\right)^{(Q-\alpha_1-\alpha_2-\alpha_3)/b} \\
&& \times {\Upsilon'(0) \Upsilon(2\alpha_1) \Upsilon(2\alpha_2) \Upsilon(2\alpha_3)\over \Upsilon(\alpha_1+\alpha_2+\alpha_3-Q)\Upsilon(\alpha_1+\alpha_2-\alpha_3)\Upsilon(\alpha_1-\alpha_2+\alpha_3)\Upsilon(-\alpha_1+\alpha_2+\alpha_3)}  \nn
\eea
  We will work in units where  $\pi \mu \gamma(b^2) b^{2-2b^2}=1$.
The four point function involving a degenerated $\phi_{12}$-field
can then be written as
\bea
F(z,\bar{z})=C_{+}\left|F^{(+)}(z)\right|^2
+C_{-}\left|F^{(-)}(z)\right|^2  \label{gg}
\eea
with
\be
C_{\pm}=
{1\over 2\pi i }  \lim_{\alpha_3\to -\ft{b}{2}}  \oint_{\gamma_{\pm}} d\alpha\, C(\alpha_1^*,\alpha_2,\alpha)C(\alpha^*,\alpha_3,\alpha_4)
 \ee
and $\gamma_{\pm}$ being a contour around $\alpha_{\pm}=\alpha_4\pm \alpha_3$.  Explicitly
\bea
C_+ &=&  \,C(\alpha_1^*,\alpha_2,\alpha_+) \nn\\
C_- &=&     \frac{C_+\,\gamma(B)\gamma(B-1)}
{\gamma(A_1)\,\gamma( A_2)\,\gamma(B-A_1)\gamma(B-A_2)}   \label{ratio}
\eea
with $A_i, B$ defined in (\ref{bfab}). One can check that $F(z,\bar{z})$ defined by (\ref{gg}) is single valued
over the whole complex plane. 

 \subsection{The CFT/gauge theory dictionary}
\label{CFTgaugedictionary}
 The gauge theory and CFT parameters are related by the dictionary (\ref{todagauge}). For $N=2$ this reduces to
\bea
\epsilon_1&=& {1\over b} \qquad  \epsilon_2=b \qquad  \epsilon=Q={1\over b}+b \qquad c=1+6 Q^2\nn\\
\alpha_1 &=& (P_0+{\bf Q} )_1=\ft{\epsilon}{2}+\ft12(\bar \mu_1-\bar \mu_2)
 ~~~~~~~~  \alpha_2 ={\kappa_0\over 2}= -\ft12 (\bar \mu_1+\bar \mu_2 )
 \nn\\
  \alpha &=&(P_1+{\bf Q} )_1= \ft{\epsilon}{2}+  a  \qquad~~~~~~~~~~~~~~~~ a=a_1=-a_2\nn\\
 \alpha_4&=& (P_2+{\bf Q} )_1= \ft{\epsilon}{2}+  \ft12( \mu_1- \mu_2)   \qquad~~\alpha_3 =  {\kappa_1\over 2}= \epsilon - \ft12 ( \mu_1+\mu_2) 
  \eea
 Using this dictionary one can check that
 \be
 Z_{\rm inst}={}_{N} F_{N-1}(^{\bf A}_{ B}\big| q)= (1-q)^{-{\alpha_2 b} } q^{-{\alpha_4  b}} \, F^{(+)}(q)
 \ee
 with the left hand side given by the gauge theory result (\ref{ZFgen}) and the right hand side by the Liouville
 conformal block  (\ref{12CFTblock1}). The term  $(1-q)^{-{\alpha_2 b} }=(1-q)^{2 \alpha_2\,\alpha_3}$
gives the  contribution of the $U(1)$ part while $q^{\alpha_4 b}=q^{\Delta_\alpha-\Delta_{\alpha_3}-\Delta_{\alpha_4}}$.\\
Analogously $Z_{{\rm inst},j=2}$ in (\ref{ZFgen2}) is given by
\be
Z_{{\rm inst},j=2}={}_{N} F_{N-1}(^{\bf A}_{ B}\big| q)= (1-q)^{-{\alpha_2 b} } q^{1-b^2+{\alpha_4  b}} \, F^{(-)}(q).
\ee

\subsection{The 't Hooft loop coefficients}
\label{sthooft}

The 't Hooft loop
coefficients $H_{\pm}$, $H_0$ are defined as \cite{Alday:2009fs}\footnote{After the exchange $\alpha_2\leftrightarrow \alpha_4$.}
 \footnotesize
\bea
H_0(a) &=&\frac{  4 \, \cos\left (  \pi b^{2} \right ) \left[  \,  \cos\left ( 2 \, \pi \,b\,  P_4  \right )  \, \cos\left ( 2 \, \pi \, b P_3 \right ) 
+\cos\left ( 2 \, \pi \, b P_1  \right )  \, \cos\left ( 2 \, \pi \,b\,P_2   \right ) \right]  }{\cos\left ( 4 \, \pi  \, b\, P \right ) -\cos\left ( 2 \, \pi {b}^{2}\right ) }\nn\\
&&
+\frac{ 4 \, \cos\left ( 2 \, \pi  \,b\, P \right )  \, \left [\cos\left ( 2 \, \pi \, b\,P_1 \right )  \, \cos\left ( 2 \, \pi \, b\, P_3 \right )
  +\cos\left ( 2 \, \pi \, b\, P_4 \right )  \, \cos\left ( 2 \, \pi \, b\,P_2 \right ) \right ] }{\cos\left ( 4 \, \pi\, b\,P \right ) -\cos\left ( 2 \, \pi \, b^{2}\right ) }
\nn\\
H_{\pm}(a) &=& -\frac{  2 \, {\pi}^{2}  \,  \Gamma\left ( 1+ 2 \,  b^2 \pm2\, b\,P      \right )  \, \Gamma\left ( b^2 \pm 2  b\, P  \right )  \, \Gamma\left ( \pm 2\,  b\, P\right )  \, \Gamma\left ( 1+b^2\pm 2\,b\, P \right )  }{  \Gamma\left ( -{b}^{2}\right )  \, \Gamma\left ( 1+
{b}^2\right )  \, \sin\left ( \pi b^{2} \right )   \,  
 \prod_{j,k=1}^2 \Gamma\left ( \frac{1}{2}+\frac{b^2}{2 }\pm b \,P \pm  \left ( -1\right )^{j} b\, P_1 \pm    \left ( -1\right )^{k} b\, P_4  \right)   }\nn\\
&&\times {1\over   \prod_{j,k=1}^2
  \Gamma\left ( \frac{1}{2}+\frac{b^2}{2}\pm b \,P \pm {\left ( -1\right ) }^{j} b\, P_3  \pm  {\left ( -1\right ) }^{k} \,b\,P_2  \right ) } \label{hhs}
\eea
\normalsize
with
\be
P_i=\alpha_i -{Q\over 2}     \qquad ~~~~~~~~~~~~P=a-{Q\over 2} 
\ee
 Specializing to $\alpha_3=-{b\over 2}$ and using the dictionary (\ref{bfab}) one can check the highly non-trivial relations
 \bea
 c_{11}&=& H_0\left(\alpha_2-{b\over 2} \right) \tilde c_1 \qquad ~~~~~~c_{12} = 2\pi H_-\left(\alpha_2+{b\over 2} \right)\tilde c_2 \,(1-q)^{A_1+A_2-B}\nn\\
 c_{22}&=&H_0\left(\alpha_2+{b\over 2}\right) \tilde c_2 \qquad ~~~~~~c_{21} = 2\pi H_+\left(\alpha_2-{b\over 2} \right) \tilde c_1 \,(1-q)^{B-A_1-A_2}
 \eea
 between the $c_{ij}$ functions defined in (\ref{cij}) and the H-functions in (\ref{hhs}). 

 \section{Instanton partition function}

 In this appendix we review the derivation of the formula (\ref{zalbe}) for the instanton partition function $Z_{Y_u Y_v}$
associated to the pair $(Y_u,Y_v)$ of Young tableaux. We refer the reader to \cite{Flume:2002az,Bruzzo:2002xf,Nakajima:2003pg} for the original references and details.

 In a D(-1)-D3 brane realization of the instanton moduli space, the tableaux $Y_u$ and $Y_v$ describe the distributions of D(-1)-instantons  around the D3-branes at positions $a_u$ and $a_v$ respectively.
The partition function (determinant) $Z_{Y_u Y_v}$ can be associated to the character (trace)
 \bea
 {\bf T}_{u  v}&=& -V_u \, V^*_v \, (1-T_1)(1-T_2)+W_u  V^*_v  +V_u  \, W^*_v \, T_1\, T_2       \label{chiy}
\eea
with
 \be
V_u = \sum_{(i,j)\in Y_{u }}   T_{a_{u }} \, T_1^{i-1} \, T_2^{j-1} \qquad
W_u =T_{a_u}=e^{i a_{u }}  \qquad
    T_{1} = e^{i \epsilon_1} \qquad  T_{2} =e^{i \epsilon_2}
\ee
 The monomials in (\ref{chiy}) coming from the
 $V_u  V^*_v$ terms are associated to  D(-1)-D(-1) strings,
  those proportional to $W_u  V^*_v$ from D3-D(-1) strings etc. Powers of $T_{\ell}$ trace the $U(1)^2\in SO(4)$
  Lorentz charges of the instanton moduli.
    Negative contributions subtract the  degrees of freedom associated to the ADHM constraints.

 \subsection{Evaluation of $T_{u  v}$ }

In the following we show that the character ${\bf T}_{u  v}$ in (\ref{chiy}) can be written as a sum of $|Y_u|+|Y_v|$ terms with no
negative contributions.
We write
\be
  V_u =  \sum_{i=1}^{k_{u ,1} } \sum_{j=1}^{\tilde{k}_{u ,i} } T_{a_{u }} T_1^{i-1} T_2^{j-1} = \sum_{j=1}^{\tilde{k}_{u ,1} } \sum_{i=1}^{k_{u ,j} } T_{a_{u }} T_1^{i-1} T_2^{j-1}
\ee
with  $(k_{u 1},k_{u  2},\ldots )$ and $(\tilde{k}_{u 1},\tilde{k}_{u  2},\ldots )$ denoting the length of the rows and columns of $Y_u $ respectively. By definition we set $k_{u  j}=0$ for $j>\tilde{k}_{u 1}$.
 One finds
\bea
V_u (1-T_2)&=&  \sum_{i=1}^{k_{u ,1} }  T_{a_{u }} T_1^{i-1} (1-T_2^{\tilde{k}_{u ,i} } )\nn\\
V^*_v(1-T_1)&=&   \sum_{j=1}^{\tilde{k}_{v,1} }  T_{-a_{v}} T_2^{1-j} ( T_1^{1-k_{v,j} }-T_1)
\eea
Plugging this into (\ref{chiy}) one finds
 \bea
 {\bf T}_{u v}&=& -V_u \, V_v^* \, (1-T_1)(1-T_2)+W_u  V^*_v  +V_u  \, W^*_v \, T_1\, T_2
     \label{chiy2} \\
  &=&   T_{a_{u v}}  \left[ \sum_{i=1}^{k_{u ,1} }  \sum_{j=1}^{\tilde{k}_{v,1} }  (T_1^i-T_1^{i-k_{v,j}})
  (T_2^{1-j} -T_2^{1-j+\tilde{k}_{u ,i}} )
  \right.\nn\\ && \left.
   +  \sum_{i=1}^{k_{v,1} } \sum_{j=1}^{\tilde{k}_{v,i} }   T_1^{1-i} T_2^{1-j}   +  \sum_{i=1}^{k_{u ,1} } \sum_{j=1}^{\tilde{k}_{u ,i} }   T_1^{i} T_2^{j}
   \right]
    \nn
\eea
 We notice that the sum over $j$ in the first term of (\ref{chiy2}) can be extended to infinity  since $k_{v,j}=0 $ for  $ j>\tilde{k}_{v,1}$. Thinking of ${\bf T}_{u v}$ as a polynomial in $T_2$, it is easy to extract the part of  ${\bf T}_{u v}$
 with positive powers in $T_2$. Writing
 ${\bf T}_{u v}=\sum_{k\in \Z} c_k T_2^k$, we define ${\bf T}^{>0}_{u v}=\sum_{k>0} c_k T_2^k$ its positive part.
 One finds
 \be
 {\bf T}^{>0}_{u v}=   \sum_{i=1}^{k_{u ,1} }  \sum_{j=1}^{\tilde{k}_{u ,i} }
  T_{a_{u v}}    T_1^{i-k_{v,j}} T_2^{1-j+\tilde{k}_{u ,i}}
 =   \sum_{s_a\in Y_{u }}   T_{a_{u v } }   \,  T_1^{-h_{v}(s) } T_2^{v_u (s)+1}
 \label{tbig0}
 \ee
with
\bea
h_{v}(s)=k_{v,j}-i  \qquad v_u (s)=\tilde{k}_{u ,i}-j
\eea
On the other hand
\be
{\bf T}_{u v}=T_1 T_2 {\bf T}^*_{vu }   \label{tabstar}
\ee
 as follows from the first line in (\ref{chiy}). Combining (\ref{tabstar}) with (\ref{tbig0}) one finds
 \bea
{\bf T}_{u v} &=& {\bf T}^{>0}_{u v}+T_1 T_2 \,({\bf T}_{vu }^{>0})^*
\eea
leading to
\be
{\bf T}_{u v}=  T_{a_{u v} }\left(  \sum_{s\in Y_{u }}     T_1^{-h_{v}(s) } T_2^{v_u (s)+1} +
 \sum_{s\in Y_{v}}    T_1^{h_{u }(s)+1 } T_2^{-v_v(s)}  \right) \label{tty0}
 \ee
Finally, writing ${\bf T}_{uv}= \sum_i e^{i \lambda_i}$ and collecting the eigenvalues $\lambda_i$ one finds for the determinant
$Z_{Y_u Y_v}=\prod_i \lambda_i^{-1}$ the formula (\ref{zalbe}).

\end{appendix}

\providecommand{\href}[2]{#2}\begingroup\raggedright\endgroup

\end{document}